\documentclass[sigconf]{acmart}
\usepackage{booktabs} 
\usepackage{natbib}
\usepackage{epsfig}
\usepackage{amssymb}
\usepackage{amsmath}
\usepackage{amsfonts}
\usepackage{breqn}
\usepackage{multirow}
\usepackage{array}
\usepackage{bm}
\usepackage{subcaption}
\usepackage{mathtools}
\usepackage{caption}
\usepackage{url}
\usepackage{float}
\usepackage{balance}
\usepackage{algorithm}
\usepackage{algorithmic}
\usepackage{bbding}
\usepackage{flushend}
\usepackage{cleveref}
\usepackage{tabularx}
\usepackage{enumitem}

\newcommand{\bul}[1]{\underline{{#1}}}

\newcommand{\xhdr}[1]{\vspace{0.8mm}\noindent{{\bf #1}}}
\newcommand{\md}{\emph{MoHR}}
\newcolumntype{R}[1]{>{\raggedleft\let\newline\\\arraybackslash\hspace{0pt}}m{#1}}
\newcolumntype{C}[1]{>{\centering\let\newline\\\arraybackslash\hspace{0pt}}m{#1}}

\setcopyright{rightsretained}
\copyrightyear{2018} 
\acmYear{2018} 
\setcopyright{acmcopyright}
\acmConference[CIKM '18]{The 27th ACM International Conference on Information and Knowledge Management}{October 22--26, 2018}{Torino, Italy}
\acmBooktitle{The 27th ACM International Conference on Information and Knowledge Management (CIKM '18), October 22--26, 2018, Torino, Italy}
\acmPrice{15.00}
\acmDOI{10.1145/3269206.3271792}
\acmISBN{978-1-4503-6014-2/18/10}

\def\S{{\bf S}}

\def\0{{\bf 0}}
\def\1{{\bf 1}}
\def\2{{\bf 2}}
\def\3{{\bf 3}}
\def\4{{\bf 4}}
\def\5{{\bf 5}}
\def\6{{\bf 6}}
\def\7{{\bf 7}}
\def\8{{\bf 8}}
\def\9{{\bf 9}}

\def\btheta{{\bm{\theta}}}

\begin{document}
\title[MoHR: Mixtures of Heterogeneous Recommenders]{Recommendation Through Mixtures\\ of Heterogeneous  Item Relationships}

\author{Wang-Cheng Kang, Mengting Wan, Julian McAuley}
\affiliation{%
  \institution{Department of Computer Science and Engineering\\University of California, San Diego\\ \{wckang,m5wan,jmcauley\}@eng.ucsd.edu}
}

\renewcommand{\shortauthors}{W.-C. Kang et al.}

\begin{abstract}
Recommender Systems have proliferated as general-purpose approaches to model a wide variety of consumer interaction data. Specific instances make use of signals ranging from user feedback
, item relationships, geographic locality, social influence (etc.). Typically, research proceeds by showing that making use of a specific signal (within a carefully designed model) allows for higher-fidelity recommendations on a particular dataset. Of course, the real situation is more nuanced, in which a combination of many signals may be at play, or favored in different proportion by individual users. Here we seek to develop a framework that is capable of combining such heterogeneous item relationships by simultaneously modeling (a) what modality of recommendation is a user likely to be susceptible to at a particular point in time; and (b) what is the best recommendation from each modality. Our method borrows ideas from mixtures-of-experts approaches as well as 
knowledge graph embeddings.
We find that our approach naturally yields more accurate recommendations than alternatives, while also providing intuitive `explanations' behind the recommendations it provides.
\end{abstract}

%
%



\maketitle

\section{Introduction}


Understanding, predicting, and recommending activities means capturing the dynamics of a wide variety of interacting forces: users' \emph{preferences} \cite{koren2009matrix,rendle2009bpr}, the \emph{context} of their behavior (e.g.~their location \cite{PACE,liu2013learning,feng2015prme}, their role in 
their
social network \cite{zhao2014leveraging,wu2017modeling,shu2018crossfire}, their dwell time~\cite{yi2014beyond,beutel2018latent}, etc.), and the \emph{relationships} between the actions themselves (e.g.~which actions tend to co-occur~\cite{koren2008factorization,kabbur2013fism}, what are their sequential dynamics \cite{rendle2010fpmc,he2016fusing,DBLP:conf/wsdm/TangW18}, etc.). \emph{Recommender Systems} are a broad class of techniques that seek to capture such interactions by modeling the complex dynamics between users, actions, and their underlying context.

\begin{figure}[t]
\includegraphics[width=\linewidth]{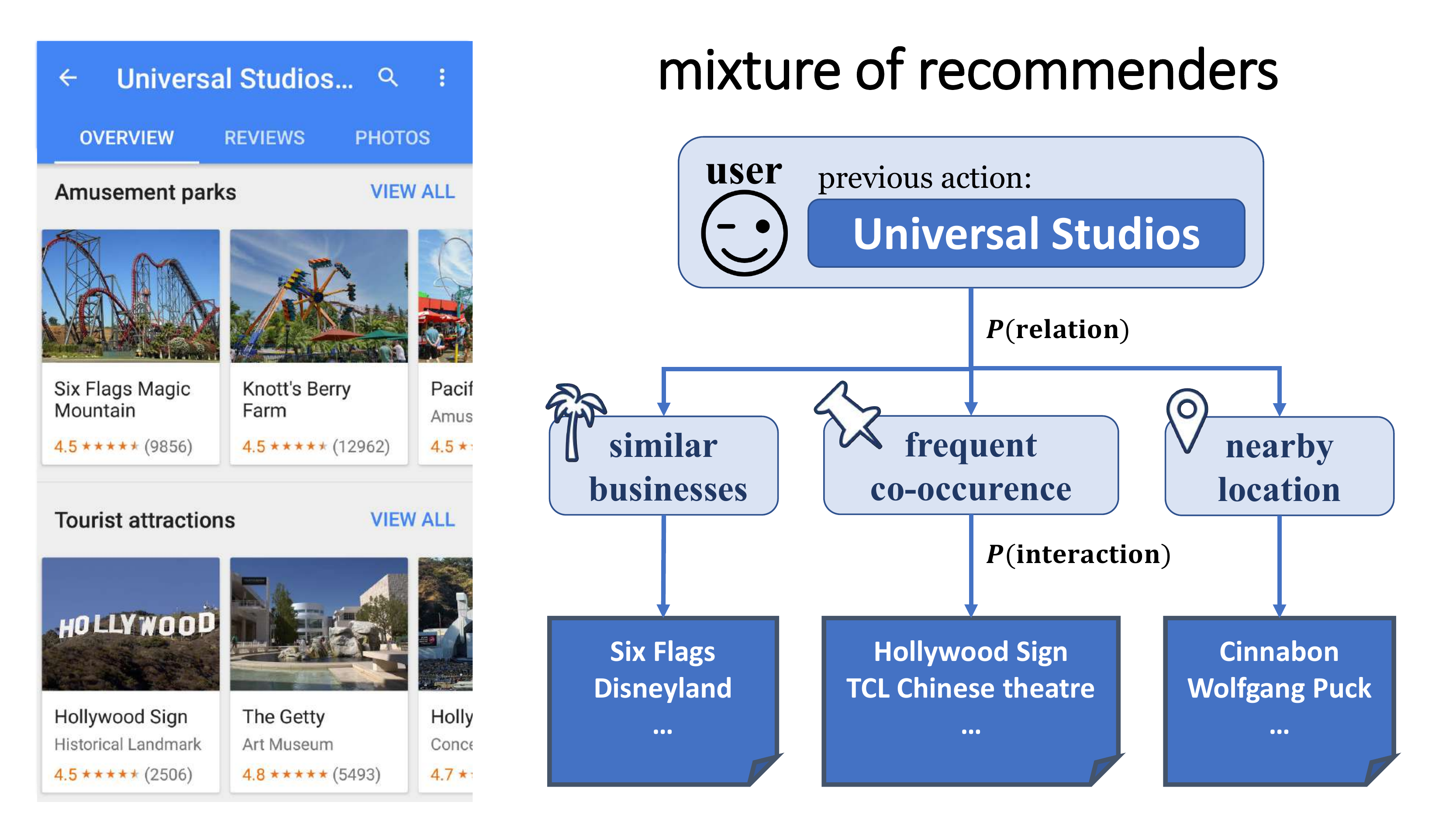}
\caption{%
A demonstration of our approach. We model recommendations as a probabilistic mixture over several heterogeneous item-to-item recommenders; when making a recommendation to a user we must then model what types of relations they're likely to adopt, as well as the relation-specific recommendations themselves. The relation types can either be manually engineered or automatically extracted from data.
\label{fig:intro}
}
\end{figure}

A dominant paradigm of research says that actions can be accurately modeled by explaining interactions between users and items, as well as interactions between items and items. User-to-item interactions might explain users' preferences or items' properties, while item-to-item interactions might describe similarity or contextual relationships between items. Examples of such models include Factorized Personalized Markov Chains (FPMC)~\cite{rendle2010fpmc} which capture user-to-item and item-to-item interactions via low-rank matrix decomposition, or more recent approaches such as TransRec~\cite{TransRec} or CKE~\cite{CKE} which seek to capture similar ideas using knowledge-graph embedding approaches.

Within these frameworks, research often proceeds by positing new types of user-to-item or item-to-item relationships that increase recommendation fidelity on a particular dataset, e.g.~ geographical similarities help POI recommendation~\cite{PACE}, ``also-viewed'' products help rating prediction~\cite{MCF}, etc. 
Each of these assumptions is typically associated with a hand-crafted model which seeks to capture the relationship in detail.

\xhdr{In this paper} we seek a more general-purpose approach to describing user-to-item and item-to-item relationships. Essentially, we note that real action sequences in any given setting simultaneously depend on several types of relationships: at different times a user might select an item based on geographical convenience, her personal preferences, the advice of her friends, or some combination of these (Figure \ref{fig:intro}); different users may weight these forms of influence in different proportions. Thus we seek a model that learns personalized mixtures over the different `reasons' why users may favor a certain action at a particular point in time.

We capture this intuition with a new model---\emph{Mixtures of Heterogeneous Recommenders} (\md{}).
Methodologically, \md{} models sequential recommendation problems in terms of a long-term preference recommender as well as a personalized, probabilistic mixture of 
heterogeneous item-to-item recommenders. 
Our method is built upon recent approaches that model recommendation in terms of translational metric embeddings~\cite{TransE,TransRec,tay2018latent,garcia2018transrev}, though in principle our model is a general framework that could be applied to any model-based
recommendation approach.

We compare \md{} against various state-of-the-art recommendation methods on multiple existing and new datasets from real applications including \emph{Amazon}, \emph{Google Local}, and \emph{Steam}. Our results show that \md{} can generate more accurate recommendations in terms of both overall and top-n ranking performance.

\section{Related Work}

\xhdr{General Recommendation.} Conventional approaches to recommendation model historical traces of user-item interactions, and at their core seek to capture users' preferences and items' properties.
Collaborative Filtering~(CF) and especially Matrix Factorization~(MF) have become 
popular underlying approaches \cite{koren2009matrix}. Due to the sparsity of explicit feedback (e.g.~rating) data, 
MF-based approaches have been proposed that make use of implicit feedback data (like clicks, check-ins, and purchases)~\cite{WRMF}. 
This idea has been extended to optimize personalized rankings of items, 
e.g.~to approximately optimize metrics such as the AUC~\cite{rendle2009bpr}. 
A recent thrust in this direction has shown that the performance of such approaches can be improved by modeling 
latent embeddings
within a metric space~\cite{CML,tay2018latent,chen2012playlist}.
Although our approach is a general purpose framework that in principle could 
be adapted to any (model-based) recommenders, we make use of metric-learning approaches due to their strong empirical performance.


\xhdr{Temporal and Sequential Recommendation.} 
The timestamps, or more simply the sequence, of users' actions provide important context to generate more accurate recommendations.
For example, TimeSVD++ sought to exploit temporal signals~\cite{koren2010temporal}, and was among the state-of-the-art methods on the \emph{Netflix} prize.
Often simply knowing the sequence of items (i.e., their ordering), and in particular the \emph{previous} action by a user, is enough to estimate 
their next action, especially in sparse datasets.
Sequential models 
often
decompose the problem into two parts: user preference modeling and sequential pattern modeling~\cite{rendle2010fpmc,feng2015prme}. Factorized Personalized Markov Chain (FPMC)~\cite{rendle2010fpmc} is a classic sequential recommendation model that fuses MF (to model user preferences) and factorized Markov Chains (to model sequential patterns).  
Recently, inspired by translational metric embeddings~\cite{TransE}, sequential recommendation models have been combined with metric embedding approaches.
 In particular, TransRec unifies the two parts by modeling each user as a translating vector from her last visited item to the next item~\cite{TransRec}. 


\xhdr{Neural Recommender Systems.} 
Various deep learning techniques have been introduced for recommendation~\cite{DBLP:journals/corr/ZhangYS17aa}. One line of work seeks to use neural networks to extract item features (e.g.~images~\cite{wang2017your,DBLP:conf/icdm/KangFWM17}, text~\cite{DBLP:conf/kdd/WangWY15,DBLP:conf/recsys/KimPOLY16}, etc.) for content-aware recommendation. Another line of work seeks to replace conventional MF. For example, NeuMF~\cite{NeuMF} estimates user preferences via Multi-Layer Perceptions~(MLPs), and AutoRec~\cite{sedhain2015autorec} predicts ratings using autoencoders. In particular, for modeling sequential user behavior, Recurrent Neural Networks~(RNNs) have been adopted to capture item transition patterns in sequences ~\cite{DBLP:journals/corr/HidasiKBT15,DBLP:conf/wsdm/JingS17,DBLP:conf/recsys/SmirnovaV17,DBLP:conf/cikm/LiRCRLM17,pei2017interacting}. In addition, CNN-based approaches have also shown competitive performance on sequential and session-based recommendation~\cite{DBLP:conf/wsdm/TangW18,DBLP:conf/recsys/TuanP17}.

\xhdr{`Relationship-aware' Recommendation.} The recommendation methods described above learn user and item embeddings from user feedback. To overcome the sparsity of user feedback, some 
methods essentially seek to `regularize' item embeddings based on item similarities or item relationships. 
For example the
POI recommendation method PACE~\cite{PACE} learns user and item representations 
while seeking to preserve item-to-item similarities based on geolocation.
\emph{Amazon} product recommendations have also benefited from the use of
``also-viewed'' products in order to improve rating prediction~\cite{MCF}; again relations among items essentially act as a form of regularizer that constrains item embeddings.
These ideas are especially useful in `cold-start' scenarios, where information from related items mitigates the sparsity of 
interactions with new items.
To exploit complex relationships, a line of work extracts item features from manually constructed meta-paths or meta-graphs in Heterogeneous Information Networks~(HIN)~\cite{yu2014personalized,zhao2017meta}.
Our method is closer to another line of work, like CKE~\cite{CKE}, which uses knowledge graph embedding techniques to automatically learn semantic item embeddings from heterogeneous item relationships.



\xhdr{Knowledge Graph Embeddings.} Originating from knowledge base domains which focus on modeling multiple, complex relationships between various entities, translating embeddings (e.g.~TransE \cite{TransE} and TransR~\cite{TransR}) have achieved state-of-the-art accuracy and scalability. 
Other than
methods like Collaborative Knowledge Base Embedding (CKE)~\cite{CKE} which use translating embeddings as regularization, several translation-based recommendation models have been proposed (e.g.~TransRec~\cite{TransRec}, LRML~\cite{tay2018latent}, TransRev~\cite{garcia2018transrev}), which show superior performance on various recommendation tasks.
Our model also adopts the translational principle to model heterogeneous interactions among users, items and relationships.

\begin{table}
\caption{Notation. \label{tb:notation}}
\begin{tabularx}{\linewidth}{lX}
\toprule
Notation&Description\\
\midrule
$\mathcal{U},\mathcal{I}$        & user and item set\\
$\mathcal{S}^u$                 & historical interaction sequence for a user $u$: $(\mathcal{S}^u_1, \mathcal{S}^u_2, ... , \mathcal{S}^u_{|\mathcal{S}^u|})$ \\
$\mathcal{R}$                    & item relationship set: $\{r_1,r_2,...,r_{|\mathcal{R}|}\}$\\
$\mathcal{\hat{R}}$                & $\mathcal{\hat{R}}=\mathcal{R}\bigcup\{r_0\}$, where $r_0$ stands for a latent relationship\\
$\mathcal{I}_{i,r}$                & item set includes all items having relation $r\in\mathcal{R}$ with item $i$\\
$K\in \mathbb{N}$                 & latent vector dimensionality\\
$\btheta_u\in\mathbb{R}^K$        & latent vector for user $u$ where $u\in\mathcal{U}$\\
$\btheta_i\in\mathbb{R}^K$        & latent vector for item $i$ where $i\in\mathcal{I}$\\
$\btheta_r\in\mathbb{R}^K$        & latent vector for relation $r$ where $r\in\mathcal{\hat{R}}$\\
$b_i\in\mathbb{R}$                & bias term for item $i$ where $i\in\mathcal{I}$\\
$b_r\in\mathbb{R}$                & bias term for relation $r$ where $r\in\mathcal{\hat{R}}$\\
$d(x,y)$                        & squared $\mathcal{L}_2$-distance between point $x$ and $y$\\
$[n]$                      	    & set of natural numbers less or equal than $n$\\
\bottomrule
\end{tabularx}
\end{table}

\section{MoHR: Mixtures of Heterogeneous Recommenders}

Our model builds on a combination of two ideas: (1) to build item-to-item recommendation approaches by making use of the translational principle (inspired by methods such as~\cite{TransE,TransRec,CKE}), followed by (2) to learn how to combine several sources of heterogeneous item-to-item relationships in order to fuse multiple `reasons' that users may follow at a particular time point. 
Following this we show how to combine these ideas within a sequential recommendation framework, and discuss parameter learning, model complexity, etc. Our notation is summarized in Table \ref{tb:notation}.

\subsection{Modeling Item-to-Item Relationships}
 
Item-to-item recommendation 
consists of identifying items related to one a user has recently interacted with, or is currently considering.
What types of items are related 
is
platform-dependent and will vary according to the domain characteristics of items. Relationships can be heterogeneous, since two items may be related because they are similar in different dimensions (e.g.~function, category, style, location, etc.) or complementary. Some work seeks to exploit one or two specific relationship(s) to improve recommendations, such as recent works that consider `also-viewed' products~\cite{MCF} or geographical similarities~\cite{PACE}. 
Extending this idea, we seek to use a general method to model such recommendation problems with heterogeneous relationships, 
improving its performance while making it easily adaptable to different domains. As we show, this approach can be applied to model a small number of hand-selected relationship types, or a large number of automatically extracted item-to-item relationships.

Specifically, for a given item $i$ and a relationship type $r$, we consider `related-item' lists containing items that exhibit relationship $r$ with item $i$
(e.g.~`also viewed' links for an item $i$).
This can equivalently be thought of as a (directed) multigraph whose edges link items via different relationship types.

Using these relationships we define a recommender to model the interaction between the three elements (two items $i$ and $i'$ linked via a relationship $r$) using a translational operation~\cite{TransE}:
\begin{equation}
R(i'|i,r)= b_{i'}-d(\btheta_i+\btheta_r, \btheta_{i'}),
\label{eq:item_predict}
\end{equation}
where $b_{i'}$ is a bias term.
This idea is drawn from knowledge graph embedding techniques where two entities (e.g.~$i=\text{Alan Turing}$ and $i'=\text{England}$) should be `close to' each other under a certain relation operation (e.g.~$r=\text{Born in}$). When used to model (e.g.) sequential relationships among items, such models straightforwardly combine traditional recommendation approaches with the translational principle~\cite{TransRec}. This idea has also been applied to several recommendation tasks and achieves 
strong
performance~\cite{TransRec,tay2018latent,garcia2018transrev}.

Similar to Bayesian Personalized Ranking~\cite{rendle2009bpr}, we minimize an objective contrasting the score of related ($i'$) versus not-related ($i^{\text{-}}$) items:
\begin{equation}
T_I = -\sum_{(i,r,i',i^{\text{-}})\in\mathcal{D}_I}\ln \sigma(R(i'|i,r)-R(i^{\text{-}}|i,r)),
\end{equation}
where 
\[\mathcal{D}_I=\{(i,r,i',i^{\text{-}})|i\in\mathcal{I}\land r\in\mathcal{R}\land i'\in\mathcal{I}_{i,r}\land i^{\text{-}}\in\mathcal{I}-\mathcal{I}_{i,r}\}.\]
The above objective encourages relevant items $i'$ to be ranked higher (i.e., larger $R(i'|i,r)$) than irrelevant items $i^{\text{-}}$ given the context of the item $i$ and relation $r$. Later we use the recommender $R(i'|i,r)$ to evaluate how closely $i$ and $i'$ are related in terms of a particular relation $r$. The recommender can also be used for inferring missing relations (as in e.g.~\cite{mcauley2015inferring}),
however doing so is not the focus of this paper and we only use the structural information as side features as in \cite{MCF,PACE,CKE}.

\subsection{Next-Relationship Prediction}

Existing item-to-user recommendation methods typically only consider item-level preferences. However, many 
applications provide multiple types of related items, e.g.~different types of co-occurrence links (also-viewed/also-bought/etc.).
Here we seek to model the problem of predicting \emph{which type of relationship a user is most likely to follow for their next interaction}. 
All items that a user selects are presumed to be related to previous items the user has interacted with, either via explicit relationships or via latent transitions. For example, when selecting a POI to visit (as in \cref{fig:intro}), a user may prefer somewhere nearby their current POI, similar to, or complementary to their current POI. 
Thus a more effective recommender system might be designed by learning to estimate which `reason' for selecting a recommendation a user will pursue at a particular point in time.

By making use of users' sequential feedback and item-to-item relationships, we can define the relevant relationships $\tau(u,k)$ given the context of the user $u$ and the $k$-th item $\mathcal{S}^u_k$ from her feedback sequence $\mathcal{S}^u$:
\begin{equation*}
\begin{split}
\tau(u,k)=\left\{
                \begin{array}{lr}
                  \{r_0\}&\forall_{r\in\mathcal{R}},\ \mathcal{S}^u_{k+1}\notin \mathcal{I}_{\mathcal{S}^u_{k},r}\\
                  \{r|\mathcal{S}^u_{k+1}\in \mathcal{I}_{\mathcal{S}^u_{k},r}\land r \in \mathcal{R}\}&\text{otherwise}\\
                \end{array}
              \right. .
\end{split}
\end{equation*}
%
The first condition in $\tau$ means that if two consecutive items share no relationship, the relevant relationship becomes a `latent' relationship $r_0$, which accounts for transitions that cannot be explained by any explicit relationship. Otherwise, 
shared relationships are relevant. Then, similarly, we define a translation-based recommender to model the interaction between the three elements:
%
\begin{equation}
R(r|u,i)= b_{r}-d(\btheta_u+\btheta_i, \btheta_r)
\label{eq:relation_predict}.
\end{equation}
Note that in contrast to \cref{eq:item_predict}---which predicts the next item under a given relationship---\cref{eq:relation_predict} predicts which \emph{relationship} will be selected following the previous item.

In particular, we define a probability function $P$ over all relationships (including $r_0$). The relevance between the relationship $r$ and context $(u,i)$ is represented by
\begin{equation}
P(r|u,i)= \frac{\exp(R(r|u,i))}{\sum_{r'\in\mathcal{\hat{R}}}\exp(R(r'|u,i))}.
\end{equation}
Again we optimize the ranking 
between relevant and irrelevant relationships by minimizing
\begin{equation}
T_R = -\sum_{(u,i,r,r^{\text{-}})\in \mathcal{D}_R}\ln \sigma(P(r|u,i)-P(r^{\text{-}}|u,i)),
\end{equation}
where 
\begin{multline*}
\mathcal{D}_R=\{(u,\mathcal{S}^u_k,r,r^{\text{-}})|u\in\mathcal{U}\land \\
k\in[|\mathcal{S}^u|-1]\land r\in \tau(u,k)\land r^{\text{-}}\in \mathcal{\hat{R}}-\tau(u,k)\}.
\end{multline*}
In addition to improving recommendations, we envisage that such a
`relation type'-level recommender can be used in personalized layout ranking,
e.g.~on the webpage of an item, we might order or organize the types of related items shown to different users according to their preferences.

\begin{figure*}[!t]
\includegraphics[width=.17\linewidth]{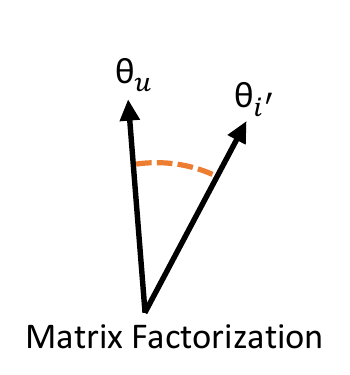}
\includegraphics[width=.17\linewidth]{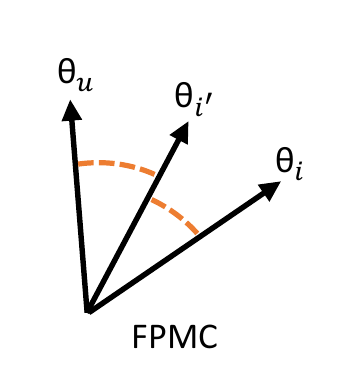}
\includegraphics[width=.17\linewidth]{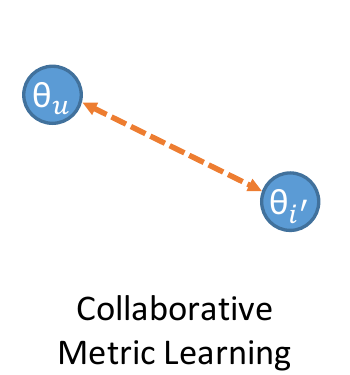}
\includegraphics[width=.17\linewidth]{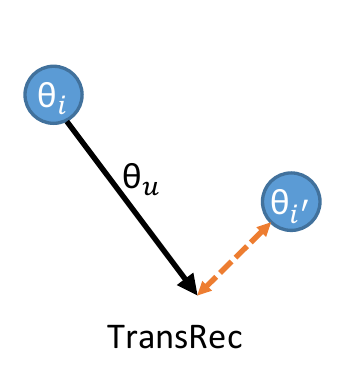}
\includegraphics[width=.17\linewidth]{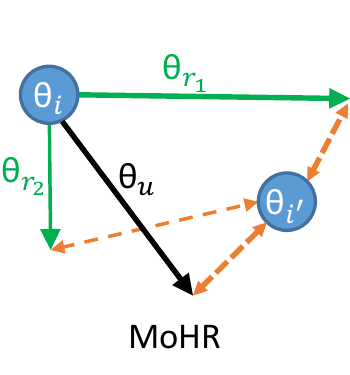}
\caption{A simplified illustration of recommendation models in existing methods. The first two models are based on inner product spaces while the following three are metric-based. The orange dashed lines indicate how a model calculates its preference score given a user $u$, her last visited item $i$ and next item $i'$. The width of lines in MoHR indicates their weight according to $P(r|u,i)$. Note that relationship-aware methods MCF~\cite{MCF} and CKE~\cite{CKE} also adopt MF as their underlying model.
}
\label{fig:models}
\end{figure*}

\subsection{Sequential Recommendation}

Like existing sequential recommenders (such as~FPMC~\cite{rendle2010fpmc} and PRME~\cite{feng2015prme}), our sequential recommender is also defined by fusing users' long-term preferences and short-term item transitions. However, rather than learning purely latent transitions, our recommendation model uses a mixture of explicit and latent item transitions. The mixture model is inspired by the `mixtures of experts' framework~\cite{moe}, which 
probabilistically mixes the outputs of different (weak) learners by 
weighting each learner according to its relevance to a given input.
In our context, each `learner' is a relationship type whose relevance is predicted given a query item. The elegance of such a framework is that it is not necessary for all learners (relationships) to accurately handle all inputs; rather, they need only be accurate \emph{for those inputs where they are predicted to be relevant}.
Here the weights on each item transition are conditioned on a user $u$ and her last visited item $i$. Specifically we define $R^*(i'|u,i)$ as:
\begin{equation*}
\begin{split}
&b_{i'}-\overbrace{d(\btheta_i+\btheta_u, \btheta_{i'})}^{\mathclap{\text{long-term preference}}}+\underbrace{P(r_0|u,i)R(i'|i,r_0)}_{\mathclap{\text{latent short-term transitions}}}+\overbrace{\sum_{r\in\mathcal{R}}P(r|u,i)R(i'|i,r)}^{\mathclap{\text{explicit short-term transitions}}},
\end{split}
\end{equation*}
Note that relation $r_0$ is a latent item relationship to capture item transitions that cannot be explained by explicit relationships. Unlike learning explicit relationships as in \cref{eq:item_predict}, $r_0$ is learned from users' sequential feedback. By unifying latent and explicit transitions, we can rewrite the recommender as:
\begin{equation}
R^*(i'|u,i)=R(i'|u,i)+\sum_{r\in\mathcal{\hat{R}}}\overbrace{P(r|u,i)}^{\mathclap{\text{probability of choosing $r$ as the next relation}}}\times \underbrace{R(i'|i,r)}_{\mathclap{\text{transition from $i$ to $i'$ using relation $r$}\ \ \ \ }}.
\label{eq:mix}
\end{equation}
Thus the
item-to-item recommenders and the next-relationship recommender
are naturally integrated into our sequential recommendation model $R^*$.

Finally, the goal of sequential recommendation is to rank the ground-truth next-item $i'$ higher than irrelevant items; the loss function we use (known as S-BPR\cite{rendle2010fpmc}) is defined as:
\begin{equation}
T_S = -\sum_{(u,i,i',i^{\text{-}})\in\mathcal{D}_S}\ln \sigma(R^*(i'|u,i)-R^*(i^{\text{-}}|u,i))\text{,}
\end{equation}
where $\mathcal{D}_S=\{(u,\mathcal{S}^u_k,\mathcal{S}^u_{k+1},i^{\text{-}})|u\in\mathcal{U}\land k\in[|\mathcal{S}^u|-1] \land i^{\text{-}}\in \mathcal{I}-\mathcal{S}^u\}$.

Figure \ref{fig:models} further illustrates
how our method compares to alternative recommendation approaches.

\subsection{Multi-Task Learning}

Different from existing methods like CKE~\cite{CKE} which introduce additional item embeddings to capture structural item information
, we use a multi-task learning framework~\cite{MTL} to jointly optimize all three tasks using shared variables within a unified translational metric space. Using shared variables can reduce the model size (see Table~\ref{tb:property} for comparison) and avoid over-fitting. Furthermore, representing different kinds of translational relationships in a unified metric space can be viewed as a form of regularization 
that fuses
different sources of data.

Specifically, we jointly learn the three tasks in a multi-task learning framework: 
\begin{equation}
\begin{split}
\min_\Theta&\quad T=T_S+\alpha T_I + \beta T_R + \lambda(\sum_{i\in\mathcal{I}} b_i^2+\sum_{r\in\mathcal{\hat{R}}} b_r^2)\\
\mathit{s.t.}&\quad \|\btheta_u\|_2\leq 1, \|\btheta_i\|_2\leq 1, \|\btheta_r\|_2\leq 1\\
&\quad \forall u\in\mathcal{U},i\in\mathcal{I}, r\in\mathcal{\hat{R}}
\end{split}
\label{eq:mtl}
\end{equation}
where $\alpha$ and $\beta$ are two hyper-parameters to control the trade-off between the main task $T_S$ and auxiliary tasks, and learnable variables $\Theta=\{\btheta_u,\btheta_i,\btheta_r,b_i,b_r\}$. We constrain the latent vectors to lie within a unit ball. This regularization doesn't push vectors toward
the origin
like $\mathcal{L}_2$ regularization, and has been shown to be effective
in both knowledge graph embedding methods~\cite{TransE,TransR} and metric-based recommendation methods~\cite{TransRec,CML,tay2018latent}. The bias terms are regularized by a square penalty with coefficient $\lambda$.

We outline the training procedure as follows:
\begin{enumerate}
\item Sample three batches from $\mathcal{D}_S$, $\mathcal{D}_I$ and $\mathcal{D}_R$, respectively
\item Update parameters using an Adam~\cite{adam} optimizer
for objective $T$ with the three batches
\item Censor the norm for all $\btheta_u,\btheta_i,\btheta_r$ by $\btheta=\btheta/\max(\|\btheta\|_2,1)$
\item Repeat this procedure until convergence
\end{enumerate}

When $\alpha=0$, we don't have semantic constraints on $R(i'|i,r)$, meaning that all relationships become latent relationships. When $\beta=0$, we don't have a prior on choosing the next relationship, meaning the model would optimize $P(r|u,i)$ only to fit sequential feedback. Typically we need to choose appropriate $\alpha>0$, $\beta>0$ to achieve satisfactory performance on the main task $T_S$. 
We discuss our hyper-parameter selection strategy in Section~\ref{sec:experiments}.

\subsection{Complexity Analysis}

The space complexity of \md{} can be calculated by the number of parameters: $(|\mathcal{U}|+|\mathcal{I}|+|\mathcal{R}|+1)*K+|\mathcal{I}|+|\mathcal{R}|+1$. Typically $|\mathcal{R}|$ is small, e.g.~$|\mathcal{R}|=2\sim101$ in the datasets we consider. Compared to CKE~\cite{CKE} which assigns each item two $K$-dimensional factors, our model saves $|I|*K$ parameters; compared to TransRec whose model size is $(|\mathcal{U}|+|\mathcal{I}|)*K+|\mathcal{I}|$, our method only adds a modest number of parameters
to model more information and tasks. The compactness of our model is mainly 
due to the
sharing of parameters across tasks.
Table~\ref{tb:property} shows an empirical comparison of the number of model parameters of various methods.

The time complexity of the training procedure is $O(NBK|\mathcal{R}|)$, where $N$ is the number of iterations and $B$ represents the batch size. At test time, the time complexity of evaluating $R^*(i'|u,i)$ is $O(K|\mathcal{R}|)$, which is larger than other methods (typically $O(K)$). However, $|\mathcal{R}|$ is typically small and our model computation (in both training and testing time) can be efficiently accelerated using multi-core CPUs or GPUs with our \emph{TensorFlow} implementation.

\subsection{Discussion of Related Methods}
\label{sec:discuss}

We examine two types of related methods and highlight the significance of our method through comparisons against them.

\xhdr{Sequential Recommendation Models.} 
We compare our method with existing next-item recommendation methods, absent any 
item relationships (i.e,~$\mathcal{R}=\{\}$). Here our recommender would become:
\[R^*_0(i'|u,i)=b_{i'}-d(\btheta_i+\btheta_u
,\btheta_{i'})-d(\btheta_i+\btheta_{r_0},\btheta_{i'}).\]

FPMC~\cite{rendle2010fpmc} combines matrix factorization with factorized first-order Markov chains to capture user preference and item transitions:
\[R_{\text{FPMC}}(i'|u,i)=\langle\btheta_u,\btheta^{(1)}_{i'}\rangle+\langle\btheta^{(2)}_i,\btheta^{(3)}_{i'}\rangle.\]
Compared to FPMC, \md{} uses a single latent vector $\btheta_i$ to reduce model size and all vectors are placed in a unified metric space rather than separate inner-product-based spaces, which empirically leads to better generalization according to~\cite{TransRec,CML}.

Recently, 
TransRec was proposed in order to introduce knowledge graph embedding techniques into recommendations~\cite{TransRec}. In TransRec, each user is modeled as a translation vector mapping her last visited item to the next item:
\[R_{\text{TransRec}}(i'|u,i)=b_{i'}-d(\btheta_i+\btheta_u,\btheta_{i'}).\]
Our method can be viewed as an extension of TransRec in that we incorporate a translation vector $r_0$ to model latent short-term item transitions.


However, our model learns a personalized mixture of explicit and latent item transitions, which is one of the main contributions in this work.

\xhdr{Relationship-Aware Recommendation Models.}
PACE~\cite{PACE} and MCF~\cite{MCF} are two recent methods that regularize item embeddings by factorizing a binary item-to-item similarity matrix $\S$, where $\S_{ij}=1$ means item $i$ and item $j$ are similar (0 otherwise). Their item embedding is used for recommendation and factorizing $\S$, which essentially is a form of multi-task learning.

Unlike the two methods above which only 
consider
homogeneous similarity, CKE~\cite{CKE} models heterogeneous item relationships with knowledge graph embedding techniques~\cite{TransE,TransR}.
To regularize the model, CKE uses the 
item embedding $\tilde{\btheta}_i$ as a Gaussian prior on the recommended item embedding $\btheta_i$. 
Note that $\btheta_i$ lies in an inner product space for Matrix Factorization while $\tilde{\btheta}_i$ is in a translational metric space for preserving item relationships. Our method uses a unified metric space with the same translational operation to represent both preferences and relationships.


The underlying recommendation models in the three methods above are simple (e.g.~inner product of a user embedding and an item embedding), while \md{}'s recommendation model directly considers users' sequential behavior and mixtures of item transitions.

\begin{table}
\begin{center}
\setlength{\tabcolsep}{3pt}
\caption{Model comparison. P: Personalized? M: Metric-based? S: Sequentially-aware? R (H-R): Models (heterogeneous) item relationship?
Model Size: The number of learnable parameters (estimated under \emph{Google Local} with $K=10$).}\label{tb:property}

\begin{tabular}{lC{.08\linewidth}C{.08\linewidth}C{.08\linewidth}C{.08\linewidth}C{.08\linewidth}c} \toprule
Property    &P&M&S&R&H-R&Model Size          \\ \midrule 
PopRec							&        &        &        &        &         &0 \\  
BPR-MF \cite{rendle2009bpr}		&\checkmark      &        &        &        &         &$99.99\%$ \\
CML \cite{CML}			&\checkmark        &\checkmark        &        &          &        &$94.41\%$ \\
FPMC \cite{rendle2010fpmc}        &\checkmark        &        &\checkmark        &        &     &$222.94\%$    \\
TransRec \cite{TransRec}    &\checkmark     &\checkmark         &\checkmark        &         &          &$99.99\%$    \\
PACE \cite{PACE}        &\checkmark      &        &        &\checkmark        &         &$150.15\%$    \\
MCF \cite{MCF}			&\checkmark      &        &        &\checkmark        &       &$150.15\%$    \\
CKE \cite{CKE}			&\checkmark      &        &        &\checkmark        &\checkmark      &$150.16\%$    \\
MoHR        &\checkmark       &\checkmark         &\checkmark        &\checkmark     &\checkmark       &$100\%$    \\ \bottomrule
\end{tabular}

\end{center}
\end{table}

\section{Experiments}
\label{sec:experiments}

\begin{table*}
\begin{center}
\setlength{\tabcolsep}{1.8pt}
\caption{Statistics of dataset after preprocessing (in ascending order of item density). }\label{tb:data}
\begin{tabular}{l|R{1.5cm}R{1.5cm}R{2cm}R{2cm}R{2.5cm}|ccc} \toprule
Dataset                &\#users \newline$|\mathcal{U}|$    &\#items \newline$|\mathcal{I}|$  &\#actions \newline $\sum_{u\in\mathcal{U}}|\mathcal{S}^u|$    &\#relationships $|\mathcal{R}|$ &\#related items \newline $\sum_{i\in\mathcal{I}}\sum_{r\in\mathcal{R}} |\mathcal{I}_{i,r}|$  &\parbox{0.08\linewidth}{\centering avg. \#actions /user}    &\parbox{0.08\linewidth}{\centering avg. \#actions /item} &\parbox{0.08\linewidth}{\centering avg. \#related items /item}  \\ \midrule 
\emph{Amazon Automotive}    &34,315        &40,287        &183,567        &4        &1,632,467        &5.35        &4.56        &40.52\\ 
\emph{Google Local}            &350,811    &505,516    &2,591,026        &101    &48,307,315        &7.39        &5.13        &95.56\\ 
\emph{Amazon Toys}            &57,617        &69,147        &410,920        &4        &3,943,494        &7.13        &5.94        &57.03\\ 
\emph{Amazon Clothing}        &184,050    &174,484    &1,068,972        &4        &2,927,534        &5.81        &6.12        &16.78\\
\emph{Amazon Beauty}        &52,204        &57,289        &394,908        &4        &2,082,502        &7.56        &6.89        &36.43\\ 
\emph{Amazon Games}            &31,013        &23,715        &287,107        &4        &1,030,990        &9.26        &12.11        &43.47\\
\emph{Steam}                &334,730    &13,047        &4,213,117        &2        &111,487        &12.59        &322.92        &8.55\\
\hline
\textbf{Total}     &\textbf{1.04M} &\textbf{0.88M} &\textbf{9.15M} &{-}  &\textbf{60.04M}    &{-}    &{-}    &{-}\\ \bottomrule
\end{tabular}
\end{center}
\end{table*}

\subsection{Datasets}

We evaluate our method on seven datasets from three large-scale real-world applications. The datasets vary significantly in domain, variability, feedback sparsity, and relationship sparsity. All the datasets we used and our code are publicly available.\footnote{\url{https://github.com/kang205/MoHR}}

\xhdr{Amazon.}\footnote{\url{https://www.amazon.com/}} A collection of datasets introduced in~\cite{mcauley2015image}, comprising large corpora of reviews as well as multiple types of related items. These data were crawled from \emph{Amazon.com} and span May 1996 to July 2014. Top-level product categories on \emph{Amazon} are treated as separate datasets.
We consider a series of large categories including `Automotive,' `Beauty,' `Clothing,' `Toys,' and `Games.' This set of data is notable for its high sparsity and variability. \emph{Amazon} surfaces four types of item relationships that we use in our model: `also viewed,' `also bought,' `bought together,' and `buy after viewing.' 

\xhdr{Google Local.}\footnote{\url{https://maps.google.com/localguides/}} A POI-based dataset~\cite{TransRec} crawled from \emph{Google Local} which contains user reviews and POIs (with multiple fine-grained categories) distributed over five continents. To construct item relationships (e.g. similar businesses), 
we first extract the top 100 categories (e.g.~restaurants, parks, attractions, etc.)~according to their frequency. Then, for each POI, based on its categories, we construct ``similar business'' relationships like ``nearby attraction,'' ``nearby park,'' etc. 
We also construct one more relationship called ``nearby popular places,'' based on each POI's geolocation and popularity. Therefore, we obtain 101 relationship types in total.

\xhdr{Steam.}\footnote{\url{http://store.steampowered.com/}} We introduce a new dataset crawled from \emph{Steam}, a large online video game distribution platform.
The dataset contains 2,567,538 users, 15,474 games and 7,793,069 English reviews spanning October 2010 to January 2018.
For each game we extract two kinds of item-to-item 
relations.
The first is ``more like this,'' based on similarity-based recommendations from \emph{Steam}.
The second are items that are
``bundled together,''
where a bundle provides a discount price for games in the same series, made by the same publisher, (etc.). Bundle data was collected from~\cite{bundle}.
The dataset also includes rich information that might be useful in future work, like user's play hours, pricing information, media score, category, developer (etc.).

We followed the same preprocessing procedure from \cite{TransRec}. For all datasets, we treat the presence of a review as implicit feedback (i.e., the user interacted with the item) and use timestamps to determine the sequence order of actions. We discard users and items with fewer than 5 related actions. For partitioning, we split the historical sequence $\mathcal{S}^u$ for each user $u$ into three parts: (1) the most recent action $\mathcal{S}^u_{|\mathcal{S}^u|}$ for testing, (2) the second most recent action $\mathcal{S}^u_{|\mathcal{S}^u|-1}$ for validation, and (3) all remaining actions for training. Hyperparameters in all cases are tuned by grid search using the validation set. Data statistics are shown in Table \ref{tb:data}. 


\begin{table*}[!ht]
\centering
\setlength{\tabcolsep}{2pt}
\caption{Ranking results on different datasets under Setting-1 (higher is better). The number of latent dimensions $K$ for all comparison methods is set to 10. The best performance in each case is underlined. The last column shows the percentage improvement of over the strongest baseline.}
\begin{tabular}{l|c|cccccccccc} \toprule
Dataset  &\parbox{1.5cm}{\centering Metric}      &\parbox{1.3cm}{\centering PopRec}  &\parbox{1.3cm}{\centering BPR-MF}    &\parbox{1.3cm}{\centering CML} &\parbox{1.3cm}{\centering FPMC} &\parbox{1.3cm}{\centering TransRec} &\parbox{1.3cm}{\centering PACE}   &\parbox{1.3cm}{\centering MCF} &\parbox{1.3cm}{\centering CKE} &\parbox{1.3cm}{\centering MoHR}    &\parbox{1.2cm}{\centering \%Improv.}  \\ \midrule
\multirow{3}{*}{\parbox{1.5cm}{\emph{Amazon Automotive}}}    
&AUC          &0.6426    &0.6313    &0.6395    &0.6414    &0.6675    &0.7233    &0.7416    &0.7341    &\bul{0.8026}    &8.2\%     \\
&HR@10        &0.3481    &0.3323    &0.3062    &0.3210    &0.3332    &0.4015    &0.4424    &0.4335    &\bul{0.5382}    &21.7\%     \\
&NDCG@10    &0.2084    &0.2003    &0.1793    &0.1981    &0.2034    &0.2371    &0.2735    &0.2607    &\bul{0.3478}    &27.2\%     \\[3pt]

\multirow{3}{*}{\parbox{1.5cm}{\emph{Google \\ Local}}}      
&AUC          &0.5811    &0.7552    &0.7676    &0.7835    &0.7915    &0.7727    &0.8560    &0.8488    &\bul{0.9330}    &9.0\%     \\
&HR@10        &0.2454    &0.5742    &0.5571    &0.5505    &0.7103    &0.5099    &0.7231    &0.7095    &\bul{0.8532}    &18.0\%     \\
&NDCG@10    &0.1380    &0.4318    &0.3995    &0.4147    &0.5400    &0.3249    &0.5484    &0.5195    &\bul{0.6091}    &11.1\%     \\[3pt]

\multirow{3}{*}{\parbox{1.5cm}{\emph{Amazon Toys}}}
&AUC          &0.6641    &0.6863    &0.7070    &0.7164    &0.7273    &0.7610    &0.7892    &0.7914    &\bul{0.8422}    &6.4\%     \\
&HR@10        &0.3601    &0.3378    &0.4015    &0.4170    &0.4474    &0.4590    &0.5277    &0.5183    &\bul{0.6061}    &14.9\%     \\
&NDCG@10    &0.2048    &0.1926    &0.2437    &0.2651    &0.2890    &0.2820    &0.3348    &0.3284    &\bul{0.4151}    &24.0\%     \\[3pt]
                                    
\multirow{3}{*}{\parbox{1.5cm}{\emph{Amazon Clothing}}} 
&AUC          &0.6609    &0.6500    &0.6527    &0.6715    &0.7034    &0.7083    &0.7529    &0.7394    &\bul{0.7882}    &4.7\%     \\
&HR@10        &0.3661    &0.3502    &0.3307    &0.3478    &0.3608    &0.3590    &0.4278    &0.4299    &\bul{0.4919}    &14.9\%     \\
&NDCG@10    &0.2166    &0.2064    &0.1904    &0.2076    &0.2111    &0.1984    &0.2601    &0.2561    &\bul{0.3015}    &16.5\%     \\[3pt]

\multirow{3}{*}{\parbox{1.5cm}{\emph{Amazon Beauty}}}
&AUC          &0.6964    &0.6767    &0.7029    &0.6874    &0.7328    &0.7638    &0.7885    &0.7805    &\bul{0.8150}    &3.4\%     \\
&HR@10        &0.4003    &0.3761    &0.4070    &0.3714    &0.4125    &0.4635    &0.5196    &0.5131    &\bul{0.5550}    &6.8\%     \\
&NDCG@10    &0.2277    &0.2164    &0.2532    &0.2107    &0.2666    &0.2820    &0.3292    &0.3245    &\bul{0.3635}    &10.4\%     \\[3pt]

\multirow{3}{*}{\parbox{1.5cm}{\emph{Amazon Games}}}
&AUC          &0.7646    &0.8107    &0.8455    &0.8523    &0.8560    &0.8632    &0.8841    &0.8849    &\bul{0.9175}    &3.4\%     \\
&HR@10        &0.4724    &0.5752    &0.6349    &0.6501    &0.6838    &0.6355    &0.7049    &0.7080    &\bul{0.7693}    &8.7\%     \\
&NDCG@10    &0.2779    &0.3249    &0.4068    &0.4576    &0.4557    &0.4044    &0.4668    &0.4582    &\bul{0.5366}    &14.5\%     \\[3pt]


\multirow{3}{*}{\emph{Steam}}
&AUC          &0.9067    &0.9233    &0.9117    &0.9219    &0.9247    &0.9012    &0.9184    &0.9115    &\bul{0.9312}    &0.7\%     \\
&HR@10        &0.7292    &0.7205    &0.7481    &0.7830    &0.7842    &0.7158    &0.7668    &0.7656    &\bul{0.7983}    &1.8\%     \\
&NDCG@10    &0.4728    &0.4655    &0.4699    &0.5297    &0.5287    &0.4663    &0.5059    &0.4829    &\bul{0.5598}    &5.7\%     \\

                                   \bottomrule
\end{tabular}
\label{tb:s1}
\end{table*}

\begin{figure*}[!h]
\centering
\begin{subfigure}[b]{0.247\textwidth}
\includegraphics[width=\linewidth]{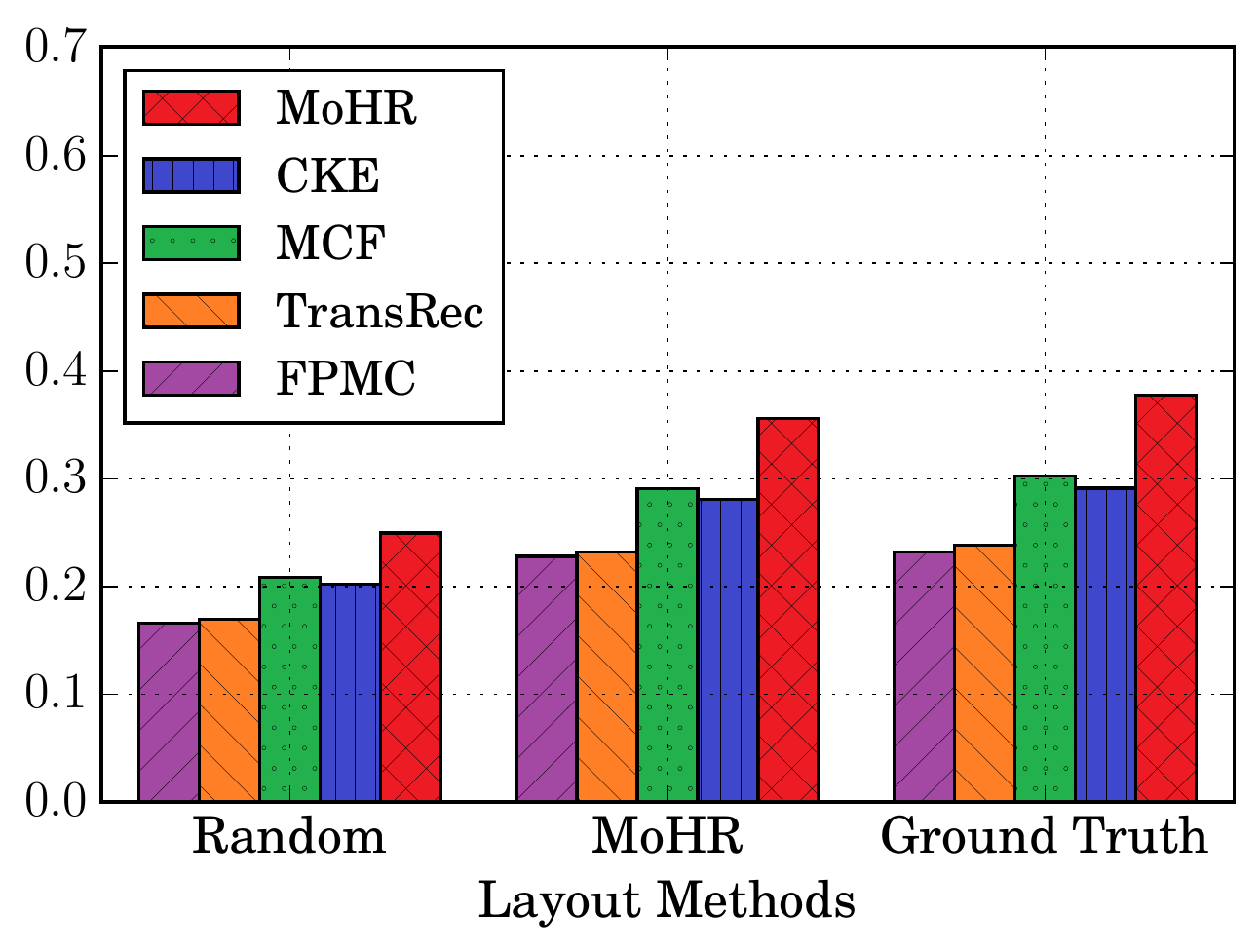}
\subcaption{Automotive}
\end{subfigure}
\begin{subfigure}[b]{0.247\textwidth}
\includegraphics[width=\linewidth]{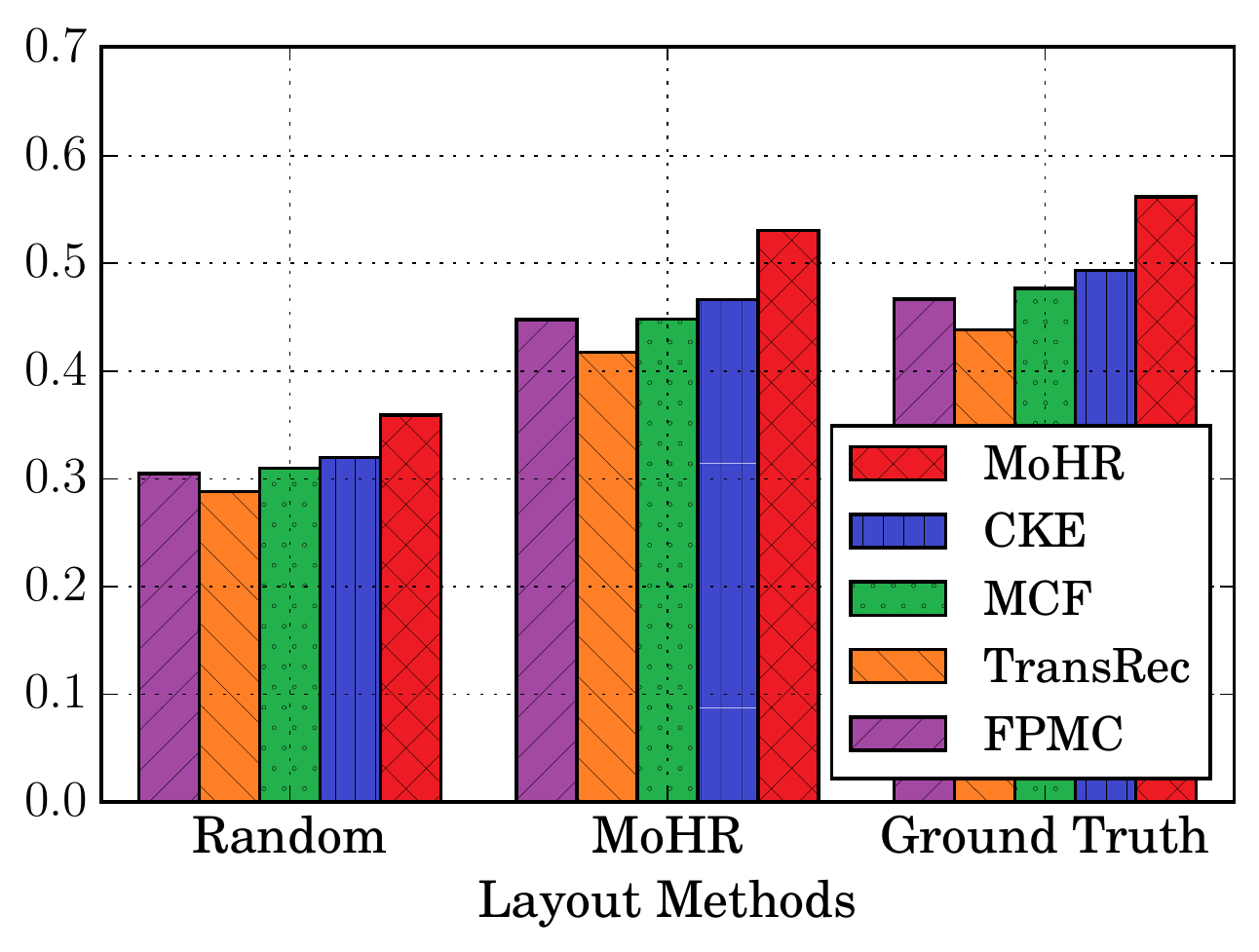}
\subcaption{Games}
\end{subfigure}
\begin{subfigure}[b]{0.247\textwidth}
\includegraphics[width=\linewidth]{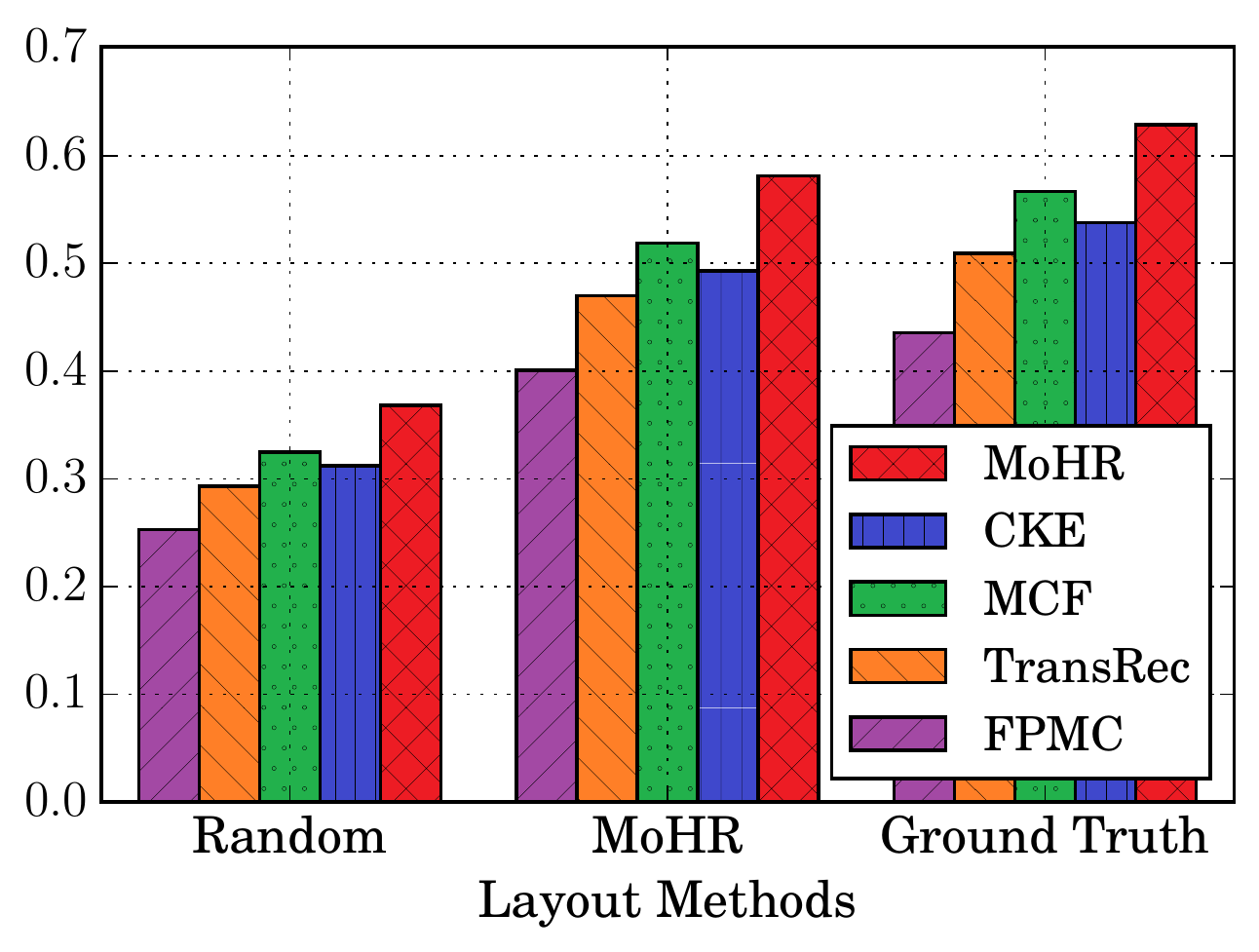}
\subcaption{Google}
\end{subfigure}
\begin{subfigure}[b]{0.247\textwidth}
\includegraphics[width=\linewidth]{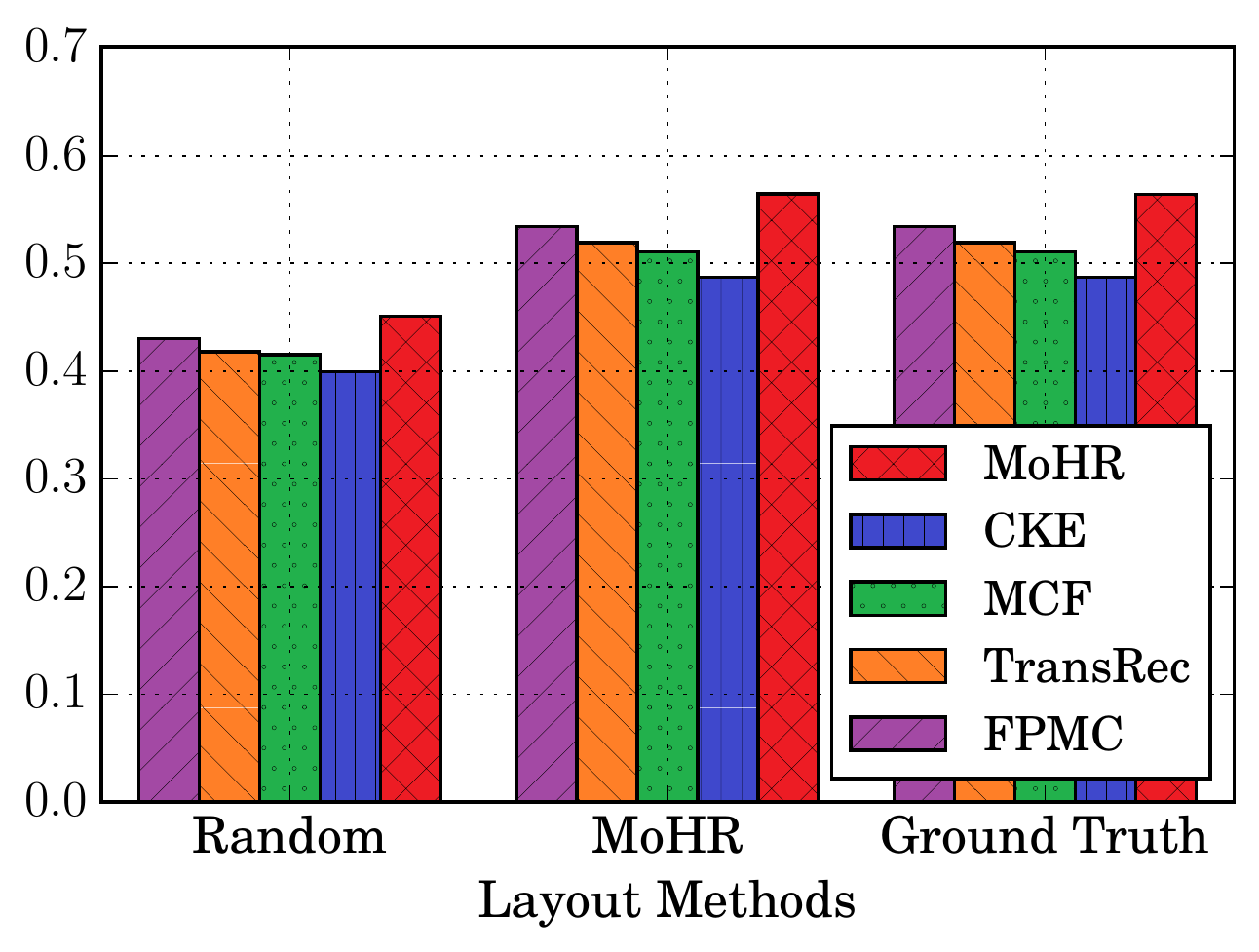}
\subcaption{Steam}
\end{subfigure}
\caption{Ranking performance (NDCG) under Setting-2 with different layout methods.}\label{fig:s2}
\end{figure*}

\subsection{Comparison Methods}

To show the effectiveness of \md{}, we include three groups of recommendation methods. The first group includes general recommendation methods which only consider user feedback without considering the sequence order of actions:
\begin{itemize}[leftmargin=*]
\item \xhdr{PopRec}: This is a simple baseline that ranks items according to their popularity (i.e., number of associated actions).

\item \xhdr{Bayesian Personalized Ranking (BPR-MF)~\cite{rendle2009bpr}}: BPR-MF is a classic method of learning personalized ranking from implicit feedback. Biased matrix factorization is used as the underlying recommender.

\item \xhdr{Collaborative Metric Learning (CML)~\cite{CML}}: A state-of-the-art collaborative filtering method that learns metric embeddings for users and items. 
\end{itemize}

The second group contains sequential recommendation methods, which 
consider the sequence of user actions:
\begin{itemize}[leftmargin=*]

\item\xhdr{Factorized Personalized Markov Chains (FPMC)~\cite{rendle2010fpmc}}: FPMC uses a  combination of matrix factorization and factorized Markov chains as its recommender, which captures users' long-term preferences as well as item-to-item transitions.

\item\xhdr{Translation-based Recommendation (TransRec)~\cite{TransRec}}: A state-of-the-art sequential recommendation method which models each user as a translation vector to capture the transition from the current item to the next item.
\end{itemize}

The final group of methods uses item relationships as regularizers: 

\begin{itemize}[leftmargin=*]

\item\xhdr{Preference and Context Embedding (PACE)~\cite{PACE}}: A neural POI recommendation method that learns user and item embeddings with regularization by preserving user-to-user and item-to-item neighborhood structure. 

\item\xhdr{Matrix Co-Factorization (MCF)~\cite{MCF}}: A recent method which showed that ``also viewed'' data helps rating prediction. It simultaneously factorizes a rating matrix and a binary item-to-item matrix based on ``also viewed'' products. 

\item\xhdr{Collaborative Knowledge base Embedding (CKE)~\cite{CKE}}: A collaborative filtering method with regularizations from visual, textual and structural item information. 
\end{itemize}

Finally, our own method,  \xhdr{Mixtures of Heterogeneous Recommenders (MoHR)},  makes use of various recommenders to capture both long-term preferences and (explicit/latent) item transitions in a unified translational metric space.


A summary of methods above is shown in Table \ref{tb:property}. For fair comparison, we implement all methods in \emph{TemsorFlow} with the Adam~\cite{adam} optimizer. All learning-based methods use BPR or S-BPR loss functions to optimize personalized rankings.
For PACE, MCF, and CKE, we do not use side information other than item relationships. For methods with homogeneous item similarities (i.e., PACE and MCF), we define two items as `neighbors' if they share at least one relationship. For CKE, we use TransE~\cite{TransE} to model item relationships.
Regularization hyper-parameters are selected from \{0,0.0001,0.001,0.01,0.1\} using our validation set. Our method can achieve satisfactory performance using $\alpha=1$, $\beta=0.1$ and  $\lambda=1e\text{-}4$ for all datasets except \emph{Steam}. Due to its high density, we use $\alpha=0.1$, $\beta=0.1$ and $\lambda=0$ for \emph{Steam}. More detailed hyper-parameter analysis is included in Section \ref{sec:hp}.

\subsection{Evaluation Metrics}



\xhdr{Setting-1.} The first is a typical sequential recommendation setting which recommends items given a user $u$ and her last visited item $i$. Here the ground-truth next item 
should be ranked as high as possible. In this setting, we report the AUC, Hit Rate@10, and NDCG@10 as in~\cite{TransRec,PACE,tay2018latent}. The AUC measures the overall ranking performance whereas HR@10 and NDCG@10 measure Top-N recommendation performance. HR@10 simply counts whether the ground-truth item is ranked among the top-10 items, while NDCG@10 is a position-aware ranking metric. 



 For top-n ranking performance metrics, to avoid heavy computation on all user-item pairs, we followed the strategy in \cite{koren2008factorization,tay2018latent,NeuMF}. For each user $u$, we randomly sample 100 irrelevant items (i.e.,~doesn't belong to $\mathcal{S}^u$), and rank these items with the ground-truth item. Based on rankings of these 101 items, HR@10 and NDCG@10 can be evaluated.

\begin{figure*}[!t]
\centering
\begin{subfigure}[b]{0.49\textwidth}
\includegraphics[width=0.5\linewidth]{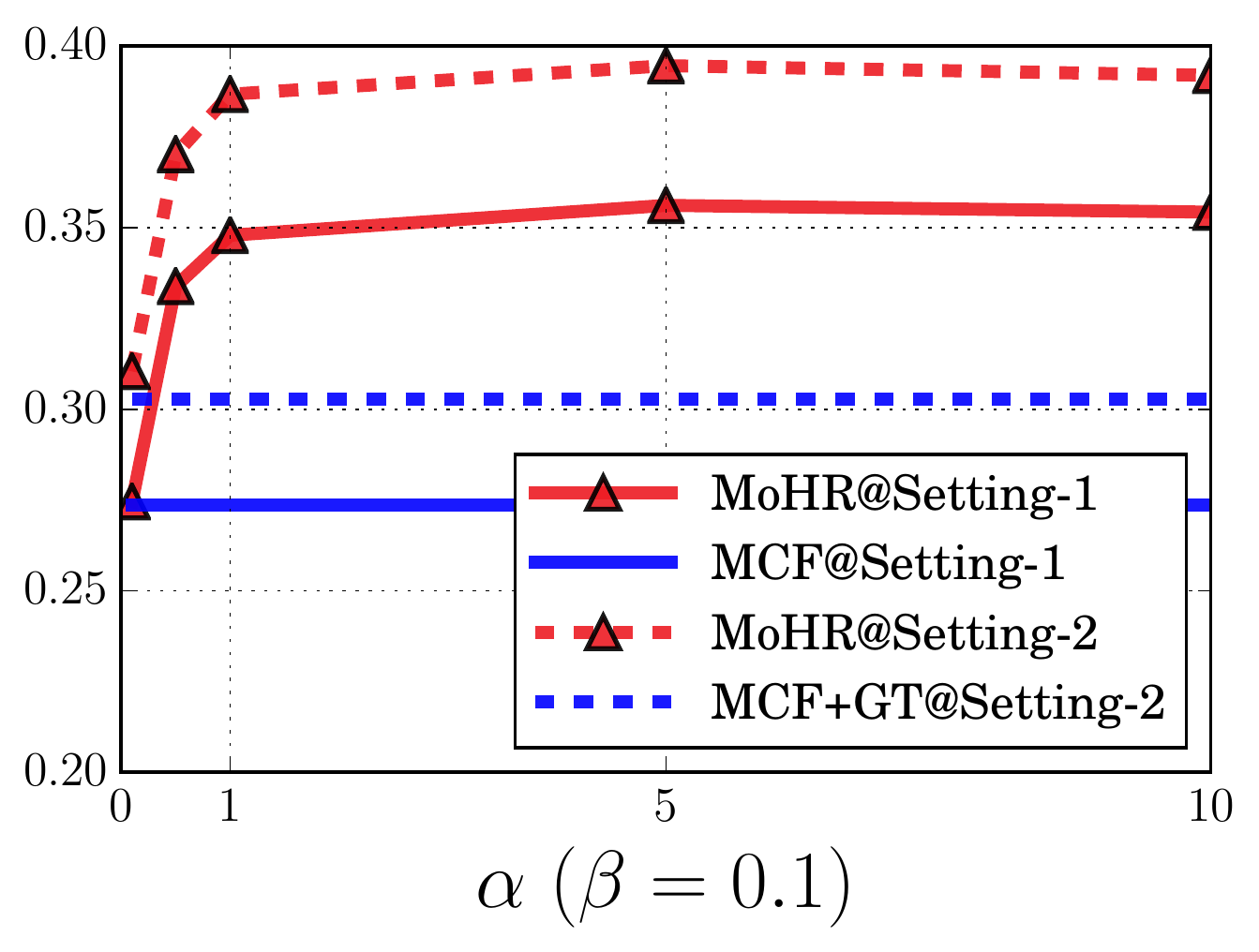}
\includegraphics[width=0.5\linewidth]{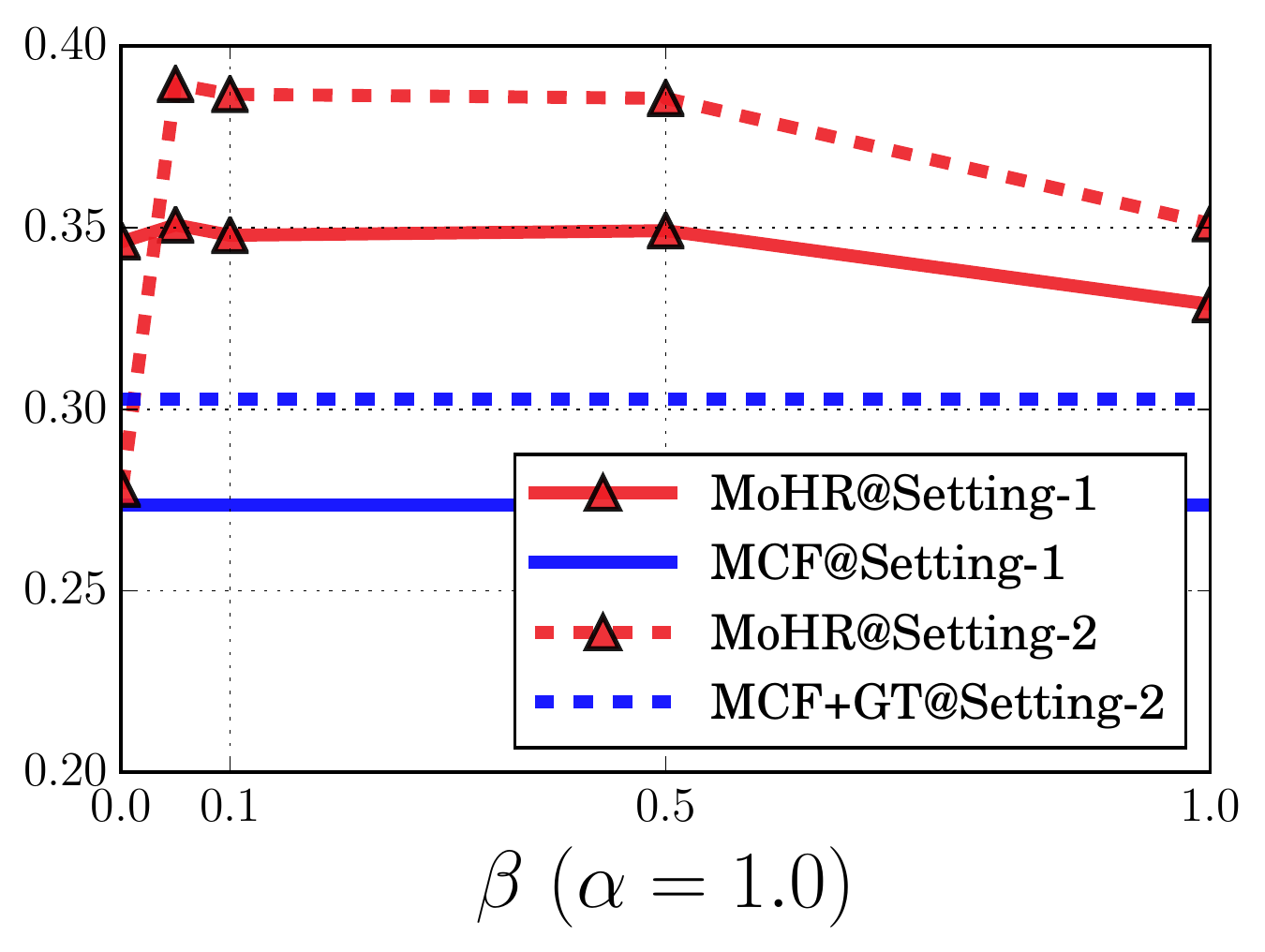}
\subcaption{Automotive}
\end{subfigure}
\begin{subfigure}[b]{0.49\textwidth}
\includegraphics[width=0.5\linewidth]{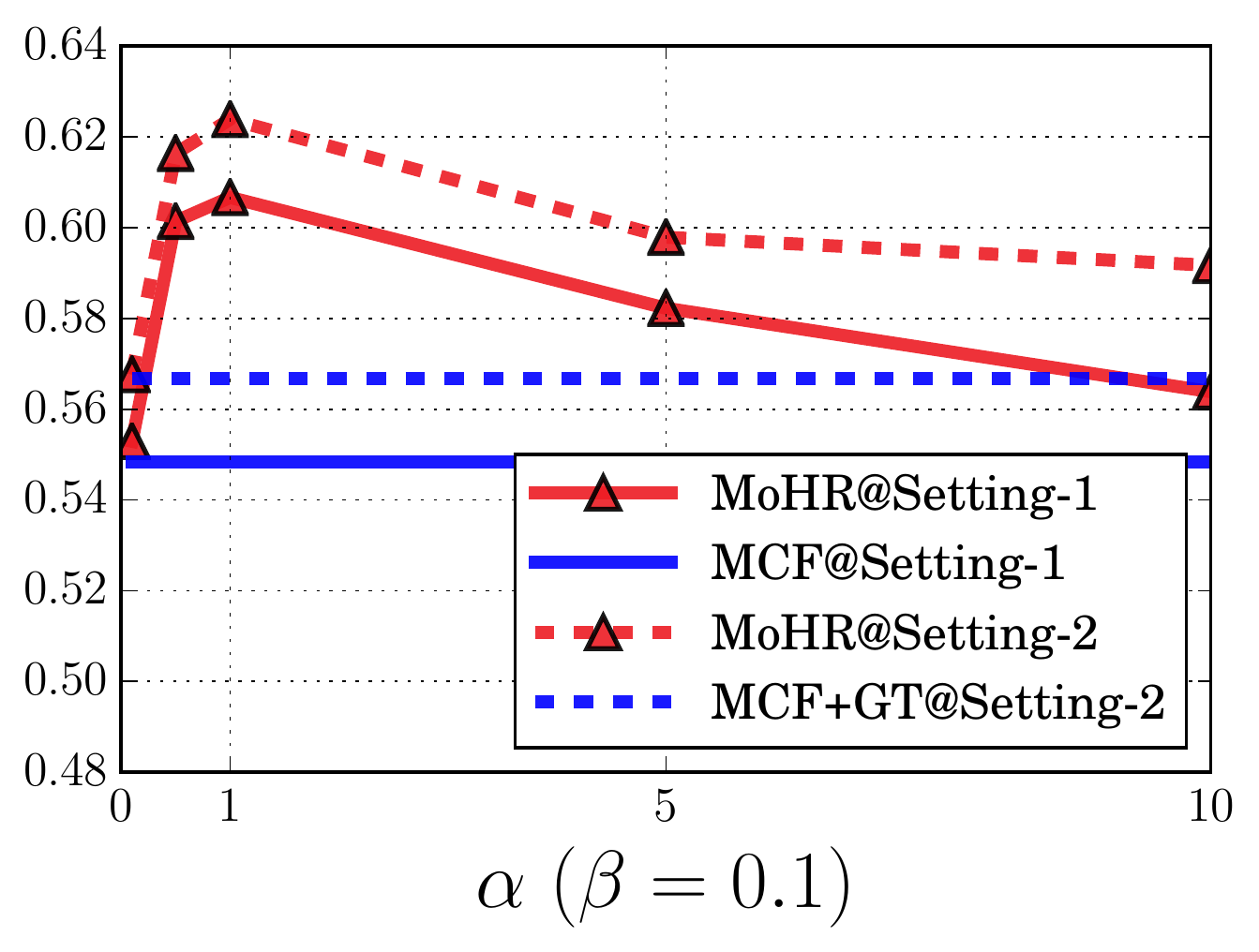}
\includegraphics[width=0.5\linewidth]{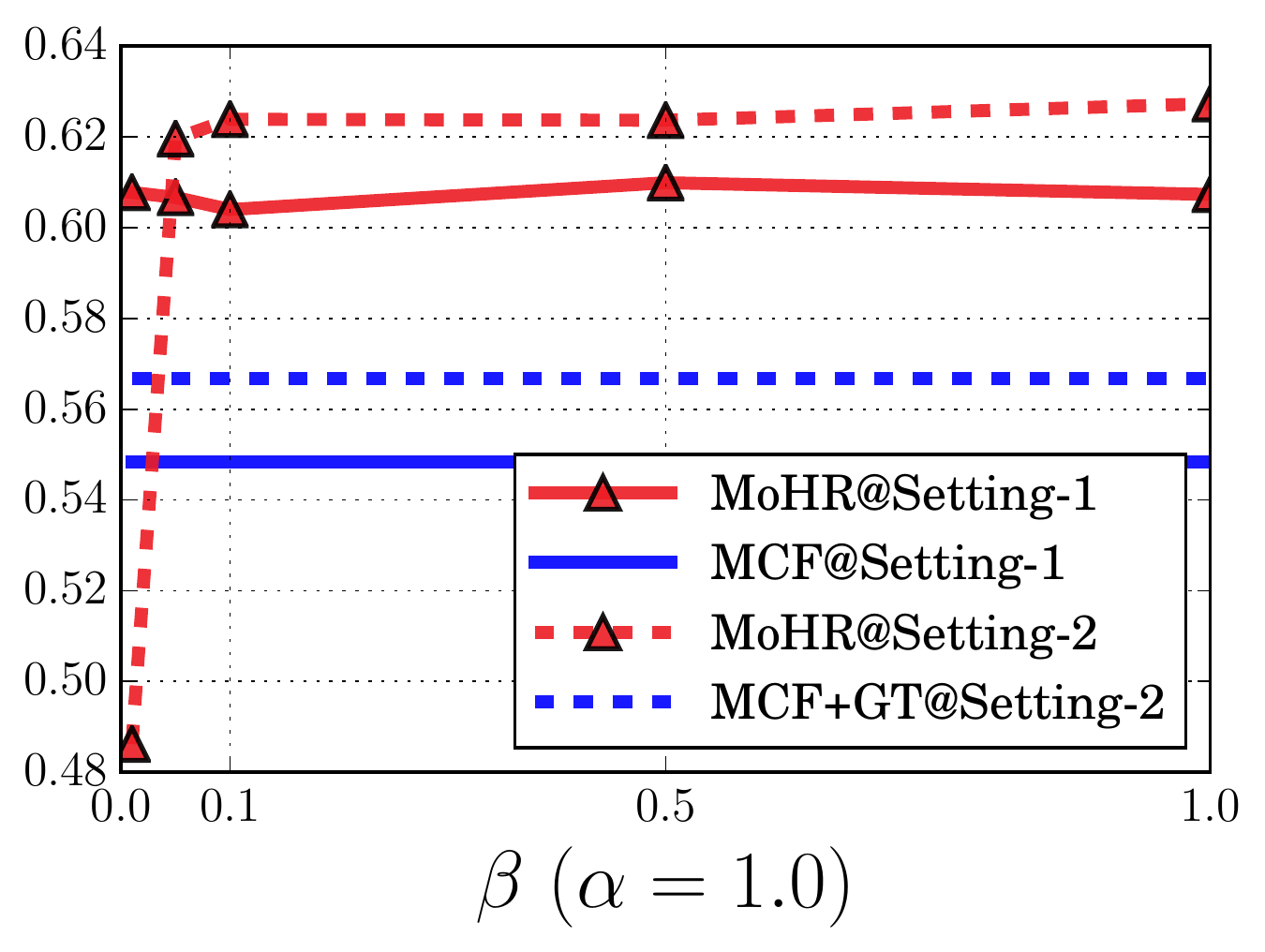}
\subcaption{Google}
\end{subfigure}

\vspace{-0.3cm}
\caption{Effect of Hyper-parameter $\alpha$ and $\beta$. Ranking performance regarding NDCG@10 and NDCG under two settings is shown in the figure respectively. The blue line indicates the strongest baseline in the corresponding dataset.}\label{fig:ab}
\vspace{-0.1cm}
\end{figure*}

\begin{figure*}
\centering
\begin{subfigure}[b]{\textwidth}
\centering
\includegraphics[width=.99\linewidth]{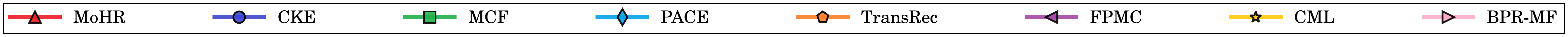}
\end{subfigure}
\begin{subfigure}[b]{0.247\textwidth}
\includegraphics[width=\linewidth]{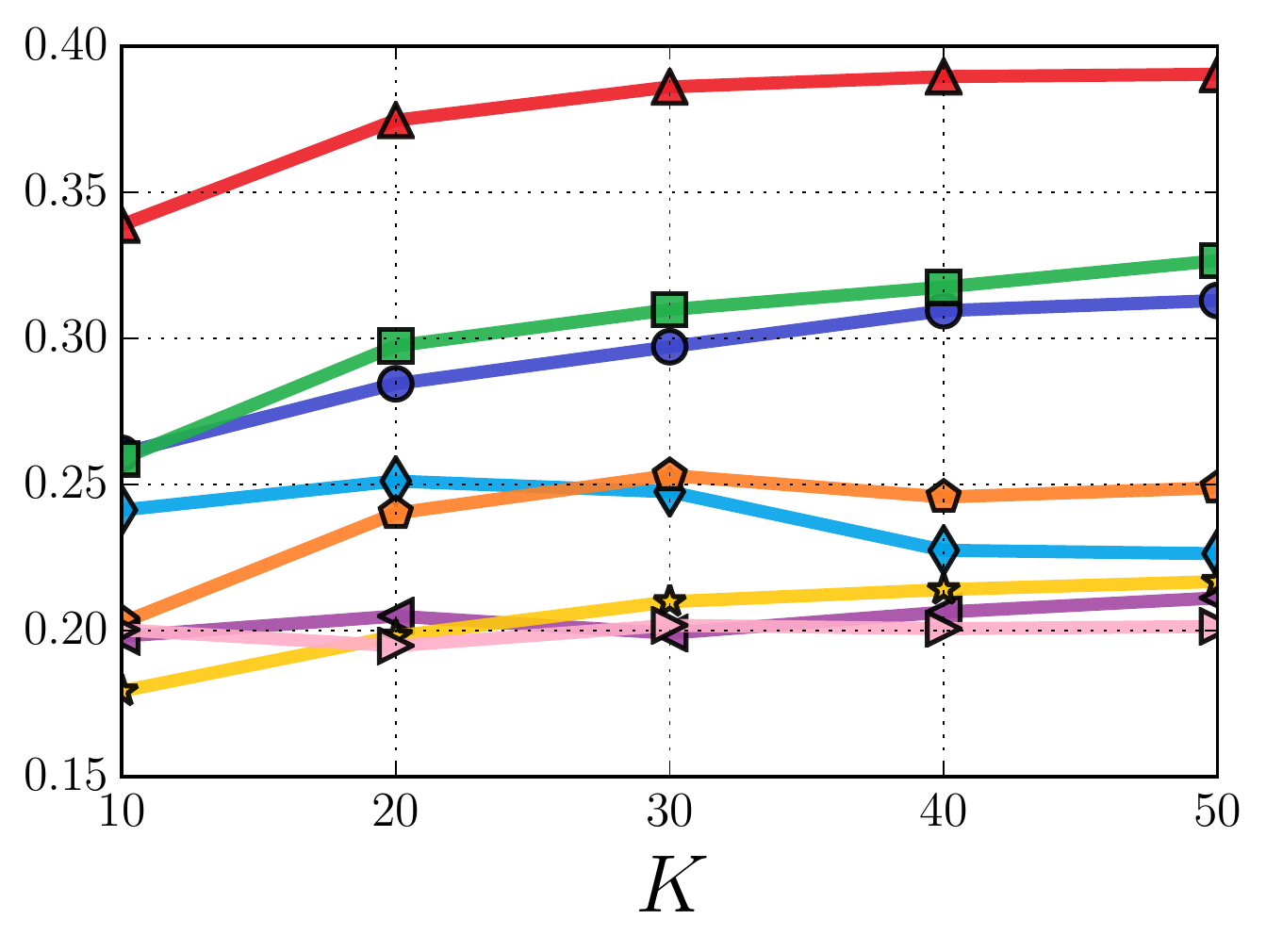}
\subcaption{Automotive}
\end{subfigure}
\begin{subfigure}[b]{0.247\textwidth}
\includegraphics[width=\linewidth]{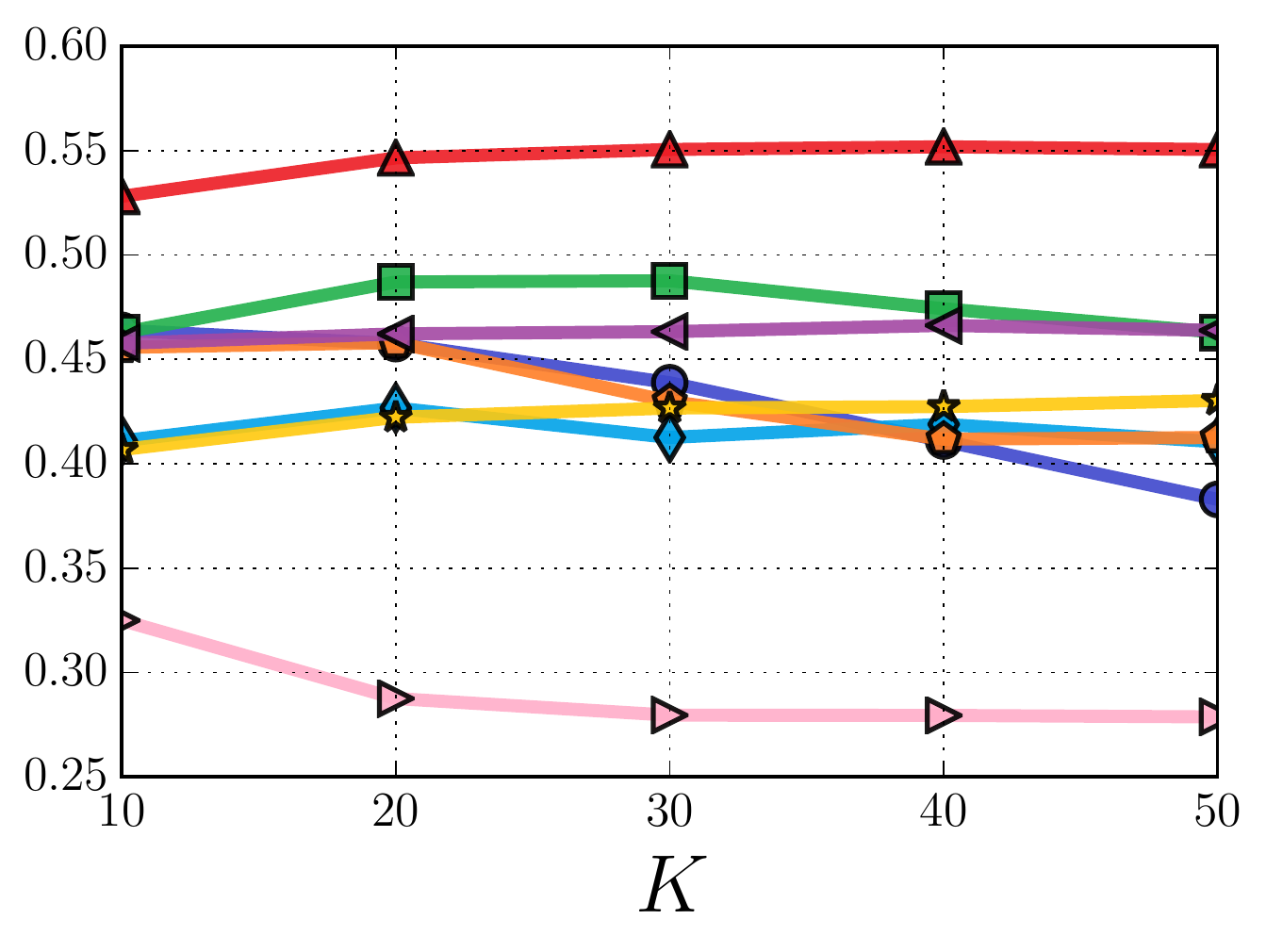}
\subcaption{Video Games}
\end{subfigure}
\begin{subfigure}[b]{0.247\textwidth}
\includegraphics[width=\linewidth]{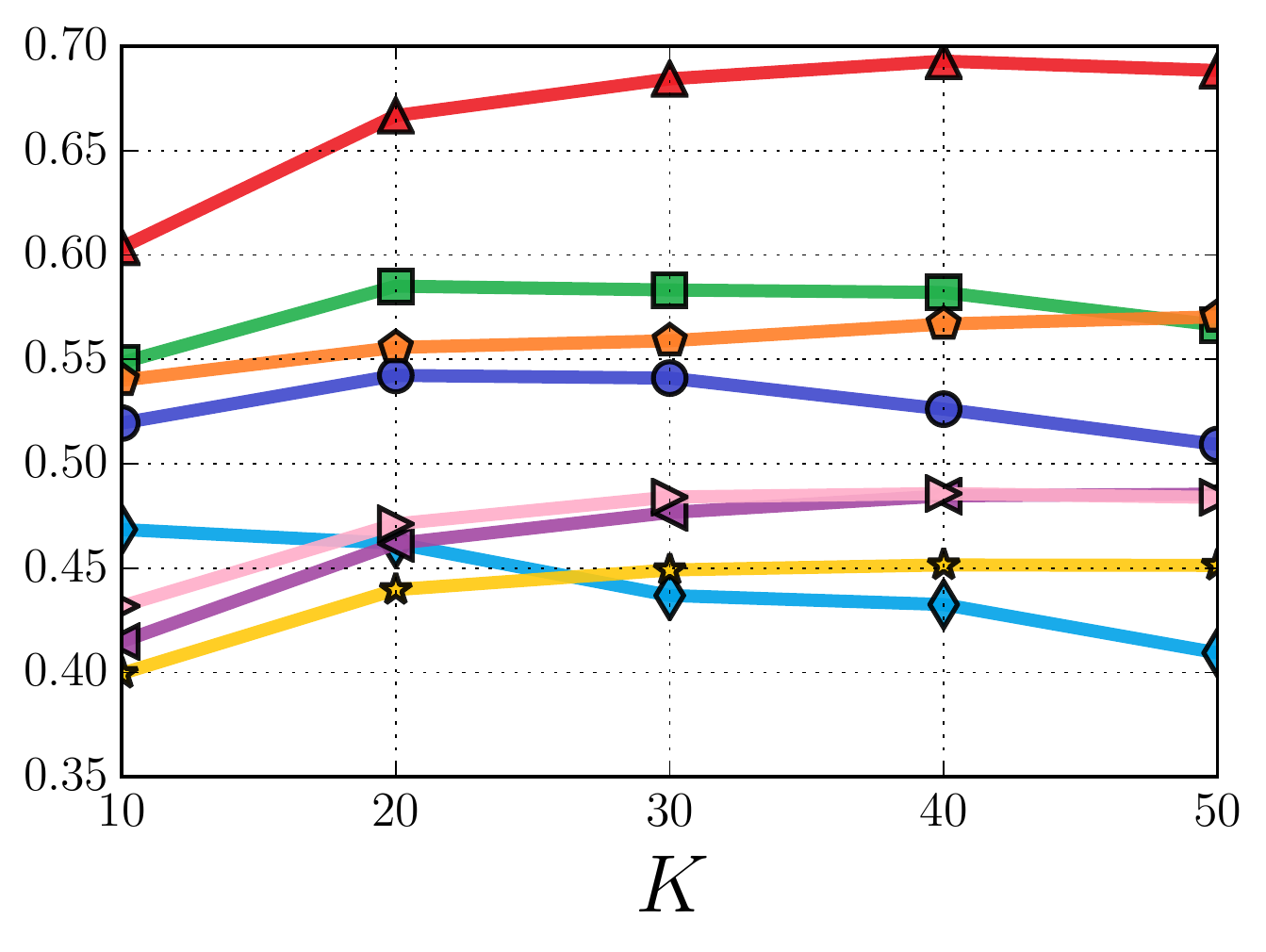}
\subcaption{Google}
\end{subfigure}
\begin{subfigure}[b]{0.247\textwidth}
\includegraphics[width=\linewidth]{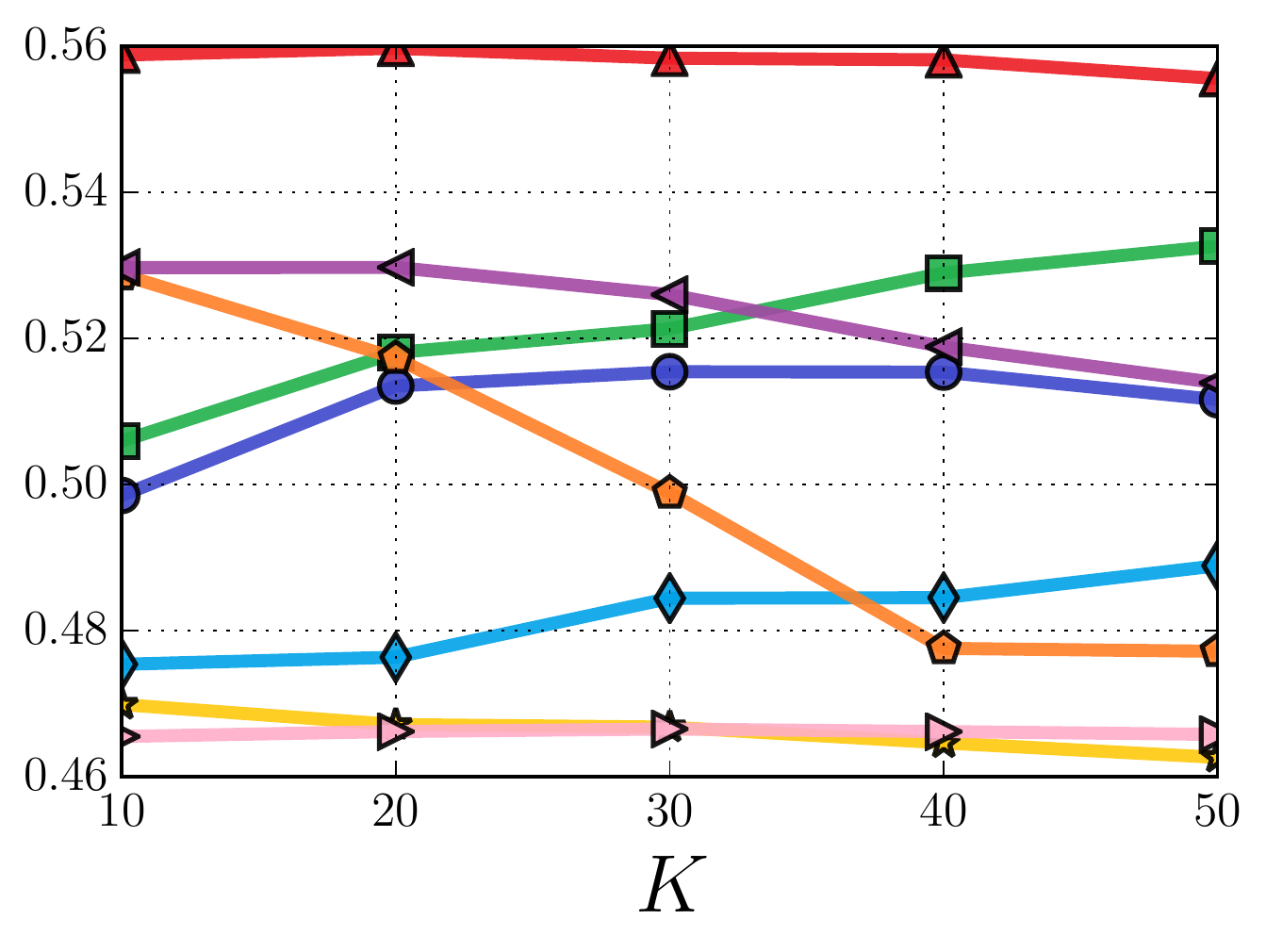}
\subcaption{Steam}
\end{subfigure}

\vspace{-0.1cm}
\caption{Effect of Hyper-parameter $K$. Ranking performance regarding NDCG@10 under Setting-1 is shown in the figure.}
\vspace{-0.2cm}
\label{fig:K}
\end{figure*}

\xhdr{Setting-2.} In the second setting, we consider a practical recommendation scenario that shows recommendations type by type (e.g.~like the interface in Figure~\ref{fig:intro} and our example in Figure~\ref{fig:example_auto}). There are two objectives: 1) relevant relationships (i.e.,~contains relevant items) should be highly ranked; and 2) within each relationship, relevant items should be highly ranked. Specifically, for a given user and her last visited item, we first rank relationships, then we display at most 10 items from each relationship. Therefore, the ultimate position of an item $i$ is decided by its ranking within the relationship as well as the ranking of the relationship to which $i$ belongs. We use NDCG to evaluate the ranking performance, which considers the positions of relevant items.


\subsection{Recommendation Performance}

Table \ref{tb:s1} shows results under the typical sequential recommendation setting. 
The number of latent dimensions $K$ is set to 10 for all experiments. Results on larger $K$ are analyzed in section \ref{sec:hp}.

We 
find our method \md{} can outperform all baselines on all datasets in terms of both overall ranking and top-n ranking metrics. Compared to sequential feedback based methods (FPMC and TransRec), the results show the important role of item-to-item relationships on understanding users' sequential behavior. Compared to methods that rely on item relationships as regularization (PACE, MCF and CKE), the 
performance
of our method shows the advantages of 
modeling item relationships and sequential signals jointly. Another observation is sequential methods FPMC and TransRec achieve better performance than relationship-ware methods on \emph{Steam}, presumably due to the high density of sequential feedback (beneficial for learning latent transitions) and sparsity of related items (insufficient for capturing item similarities) in \emph{Steam}.

For Setting-2, a recommendation method should decide the order in which to show relationships as well as the rank of items from each relationship. All methods are capable of ranking items for different users. However, only our method \md{} models the next-relationship problem and can give a prediction of relationship ranking. Figure \ref{fig:s2} shows the results of our method and four strong baselines under three relationship ranking methods: random, \md{}'s relationship ranking prediction (i.e., $P(r|u,i)$) and the ground-truth relationship ranking (i.e., relationships that contain ground-truth items are ranked higher than others). We can see \md{} can outperform the baselines under all three relationship ranking methods, and \md{} (with its own relationship ranking) can beat 
baselines with the ground-truth ranking. This shows the effectiveness of our method on both relationship and item ranking.


\subsection{Effect of Hyper-parameters}\label{sec:hp}

We empirically analyze the effect of the hyper-parameters in our model. Figure \ref{fig:ab} shows the influence of $\alpha$, $\beta$ under the two settings. When $\alpha\to0$, item vectors $\btheta_i$ and relationship vectors $\btheta_r$ are free to fit sequential feedback without the constraint from relationships. As $\alpha$ increases, the performance on both settings improves and then degrades, which is similar to the effect of regularization. $\beta$ (the hyper-parameter controlling the prior of choosing the next relationship) doesn't improve the performance significantly on Setting-1. However, when $\beta\to0$, the model is not aware of how users choose their next relationship, which leads to poor performance on 
Setting-2, due to poor relationship ranking. Typically when $\alpha=1$ and $\beta=0.1$, the model can achieve satisfactory performance on both settings. For the latent dimensionality $K$, Figure \ref{fig:K} shows the recommendation performance of all the baselines with K varying from $10$ to $50$. We can find that our method outperforms baselines across the full spectrum of values.

\subsection{Ablation Study}

We analyze the effect of each component in \emph{MoHR} by ablation study. First, recall that \emph{MoHR}'s predictor $R^*$ (in eq. (\ref{eq:mix})) contains two terms for long-term preference and mixture of short-term transitions (respectively). If the second term (for the mixture) is removed, then the predictor is equivalent to TransRec's. Hence we analyze the effect of the mixture term by comparing the performance of using/not using the mixture. Second, our optimization problem contains two auxiliary tasks ($T_I$ and $T_R$ in eq. (\ref{eq:mtl})). Without the two tasks, our model is not able to utilize explicit item relationships. By enumerating all the options, we induce four variants in total, and their recommendation performance (NDCG@10) is shown in Table \ref{tb:abla}.

We can see that multi task learning can significantly boost the performance on all datasets, implying the importance of using explicit item transitions for next item recommendation. Also, using the mixture of short-term transitions consistently improves performance when multi task learning is employed. For single task cases where all transitions are latent, using the mixture can boost performance on all datasets except \emph{Google}. Overall, the full model of \emph{MoHR} (multi task+mixture) achieves the best performance.

\begin{table}[h]
\centering
\caption{Performance using different components of MoHR \label{exp:deep}}
\begin{tabular}{lcccc}
\toprule
Component					& \emph{Auto}		&	\emph{Video}	&	\emph{Google}	&	\emph{Steam}\\ \midrule                               
Single Task  				&	0.1805	&	0.4617	&	0.4371	&	0.5276 \\
Single Task + Mixture 		&	0.1864	&	0.4626	&	0.4069	& 	0.5360 \\                      
Multi Tasks 					&	0.3304 	&	0.5214	&	0.6002	&	0.5588 \\
Multi Tasks + Mixture		&	0.3478	&	0.5366	&	0.6091	&	0.5598 \\
\bottomrule
\end{tabular}
\label{tb:abla}
\end{table}

\subsection{Comparison to Deep Learning Approaches}

We also compare our methods against deep learning based methods that seek to capture sequential patterns for recommendation. Specifically, two baselines are : 1) \textbf{GRU4Rec}~\cite{DBLP:journals/corr/HidasiKBT15} is a seminal method that uses RNNs for session-based recommendation. For comparison we treat each user's feedback sequence as a session; 2) \textbf{GRU4Rec~(revised)}~\cite{DBLP:journals/corr/HidasiK17} is an improved version which adopts a different loss function and sampling strategy; 3) \textbf{Convolutional Sequence Embedding (Caser)}~\cite{DBLP:conf/wsdm/TangW18} is a recent method for sequential recommendation, which treats embeddings of preceding items as an `image' and extracts sequential features by applying convolutional operations. We use the code from the corresponding authors, and conduct experiments on a workstation with a single GTX 1080 Ti GPU.

Table \ref{exp:deep} shows the Top-N ranking performance (NDCG@10) of baselines and our method \emph{MoHR} on four datasets. \emph{MoHR} outperforms all baselines significantly on all datasets except \emph{Steam}. GRU4Rec~(revised) shows the best performance on \emph{Steam}, while \emph{MoHR} is slightly worse. Presumably this is because \emph{Steam} is a relatively dense dataset, where CNN/RNN-based methods can learn higher order sequential patterns while \emph{MoHR} only considers the last visited item. We can also see the gap between Caser and \emph{MoHR} on \emph{Steam} is smaller than that on other datasets, which could also be attributed to density. Though deep learning based methods are computationally expensive and don't tend to perform well on sparse datasets, 
in the future we hope to incorporate them into \emph{MoHR} to capture higher-order item transitions
while alleviating data scarcity problems by considering explicit item relationships.
\begin{table}[h]
\centering
\caption{Comparison to CNN/RNN-based Methods \label{exp:deep}}
\begin{tabular}{lcccc}
\toprule
Method															& \emph{Auto}		&	\emph{Video}	&	\emph{Google}	&	\emph{Steam}\\ \midrule                               
GRU4Rec (original)~\cite{DBLP:journals/corr/HidasiKBT15} 		&	0.0848	&	 0.2250	&	0.4166	& 	0.3521 \\
GRU4Rec (revised)~\cite{DBLP:journals/corr/HidasiK17}			&	0.1246	&	0.3132	&	0.4032	&	\bf{0.5676}\\
Caser~\cite{DBLP:conf/wsdm/TangW18} 							&	0.2518	& 0.3075	&	0.2244	&	0.5141  \\                            
MoHR		&	\bf{0.3478}	&	\bf{0.5366}	&	\bf{0.6091}	&	0.5598 \\
\bottomrule
\end{tabular}
\end{table}
\subsection{Qualitative Examples}

\begin{figure}[htb]
\vspace{-0.3cm}
\includegraphics[width=.9\linewidth]{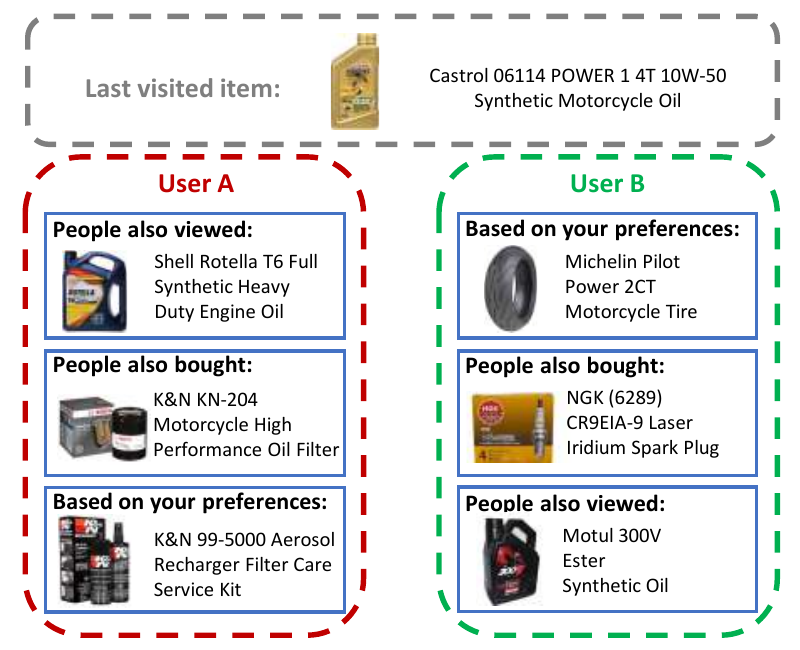}
\vspace{-0.1cm}
\caption{A recommendation example for two users with the same last visited item. For each type of recommendation, we display the top result. The order of relationships is decided by $P(r|u,i)$. This shows how our model generates different layouts according to user context.}
\label{fig:example_auto}
\end{figure}

Figure \ref{fig:example_auto} illustrates a practical recommendation scenario using our method \md{} on the \emph{Amazon Automotive} dataset. We can see that \md{} can generate a layout and recommendations according to the context $(u,i)$. This may also provide a way to interpret users' intentions behind their behavior. For example, User A in the figure may want to try another brand of oil whereas User B may seek complementary accessories.

Recall that we use $r_0$ to capture latent transitions (unexplained by explicit relationships) from sequential feedback. To show the latent transitions learned by transition vector $\btheta_{r_0}$,  We train a model without bias terms (for better illustration) on \emph{Google Local}. Table \ref{tb:google} shows consecutive transitions starting from various popular POIs. Unlike explicit relationships which only represent ``nearby businesses'' and ``similar businesses,'' the latent relationship can capture richer semantics and long-range transitions from sequential feedback.

\begin{table}[htb]
\centering
\setlength{\tabcolsep}{0pt}
\footnotesize
\small
\caption{Transition examples using the latent relationship $r_0$. Place $i$'s next place is the nearest neighbor of $\btheta_i+\btheta_{r_0}$. We can see $r_0$ is able to capture meaningful transitions with various semantics and ranges.}
\label{tb:google}
\begin{tabular}{l}
\toprule

Universal Studios Hollywood $\rightarrow$ Hollywood Sign $\rightarrow$ LAX Airport
\\
Universal Studios Orlando $\rightarrow$ Disney World Resort $\rightarrow$ Florida Mall 
\\
Universal Studios Japan $\rightarrow$ Yodobashi Akiba $\rightarrow$ Edo-Tokyo Museum 
\\
JFK Airport (New York) $\rightarrow$ Statue of Liberty $\rightarrow$ Empire State Building
\\
Heathrow Airport (London)$\rightarrow$ London Bridge $\rightarrow$ Croke Park (Ireland) 
\\
Hong Kong Airport $\rightarrow$ The Peninsula Hong Kong 
$\rightarrow$ Hong Kong Park
\\
Death Valley National Park $\rightarrow$ Circus Circus (Las Vegas) 
$\rightarrow$ Yellowstone
\\
Louvre Museum $\rightarrow$ Eiffel Tower $\rightarrow$ Charles de Gaulle Airport (Paris) 
\\

\bottomrule
\end{tabular}
\setlength{\tabcolsep}{6pt}
\end{table}


\section{Conclusion}

In this work we present a sequential recommendation method \md{}, which learns a personalized mixture of heterogeneous (explicit/latent) item-to-item recommenders. We represent all parameters in a unified metric space and adopt translational operations to model their interactions. Multi-task learning is employed to jointly learn the embeddings across the three tasks. Extensive quantitative results on large-scale datasets from various real-world applications demonstrate the superiority of our method regarding both overall and Top-N recommendation performance.



\bibliographystyle{ACM-Reference-Format}
\bibliography{acmart}


\begin{thebibliography}{48}


\ifx \showCODEN    \undefined \def \showCODEN     #1{\unskip}     \fi
\ifx \showDOI      \undefined \def \showDOI       #1{#1}\fi
\ifx \showISBNx    \undefined \def \showISBNx     #1{\unskip}     \fi
\ifx \showISBNxiii \undefined \def \showISBNxiii  #1{\unskip}     \fi
\ifx \showISSN     \undefined \def \showISSN      #1{\unskip}     \fi
\ifx \showLCCN     \undefined \def \showLCCN      #1{\unskip}     \fi
\ifx \shownote     \undefined \def \shownote      #1{#1}          \fi
\ifx \showarticletitle \undefined \def \showarticletitle #1{#1}   \fi
\ifx \showURL      \undefined \def \showURL       {\relax}        \fi
\providecommand\bibfield[2]{#2}
\providecommand\bibinfo[2]{#2}
\providecommand\natexlab[1]{#1}
\providecommand\showeprint[2][]{arXiv:#2}

\bibitem[\protect\citeauthoryear{Beutel, Covington, Jain, Xu, Li, Gatto, and
  Chi}{Beutel et~al\mbox{.}}{2018}]%
        {beutel2018latent}
\bibfield{author}{\bibinfo{person}{Alex Beutel}, \bibinfo{person}{Paul
  Covington}, \bibinfo{person}{Sagar Jain}, \bibinfo{person}{Can Xu},
  \bibinfo{person}{Jia Li}, \bibinfo{person}{Vince Gatto}, {and}
  \bibinfo{person}{Ed~H Chi}.} \bibinfo{year}{2018}\natexlab{}.
\newblock \showarticletitle{Latent Cross: Making Use of Context in Recurrent
  Recommender Systems}. In \bibinfo{booktitle}{\emph{WSDM'18}}.
\newblock


\bibitem[\protect\citeauthoryear{Bordes, Usunier, Garcia-Duran, Weston, and
  Yakhnenko}{Bordes et~al\mbox{.}}{2013}]%
        {TransE}
\bibfield{author}{\bibinfo{person}{Antoine Bordes}, \bibinfo{person}{Nicolas
  Usunier}, \bibinfo{person}{Alberto Garcia-Duran}, \bibinfo{person}{Jason
  Weston}, {and} \bibinfo{person}{Oksana Yakhnenko}.}
  \bibinfo{year}{2013}\natexlab{}.
\newblock \showarticletitle{Translating embeddings for modeling
  multi-relational data}. In \bibinfo{booktitle}{\emph{NIPS'13}}.
\newblock


\bibitem[\protect\citeauthoryear{Caruana}{Caruana}{1997}]%
        {MTL}
\bibfield{author}{\bibinfo{person}{Rich Caruana}.}
  \bibinfo{year}{1997}\natexlab{}.
\newblock \showarticletitle{Multitask Learning}.
\newblock \bibinfo{journal}{\emph{Machine Learning}} (\bibinfo{year}{1997}).
\newblock


\bibitem[\protect\citeauthoryear{Chen, Moore, Turnbull, and Joachims}{Chen
  et~al\mbox{.}}{2012}]%
        {chen2012playlist}
\bibfield{author}{\bibinfo{person}{Shuo Chen}, \bibinfo{person}{Josh~L Moore},
  \bibinfo{person}{Douglas Turnbull}, {and} \bibinfo{person}{Thorsten
  Joachims}.} \bibinfo{year}{2012}\natexlab{}.
\newblock \showarticletitle{Playlist prediction via metric embedding}. In
  \bibinfo{booktitle}{\emph{KDD'12}}.
\newblock


\bibitem[\protect\citeauthoryear{Feng, Li, Zeng, Cong, Chee, and Yuan}{Feng
  et~al\mbox{.}}{2015}]%
        {feng2015prme}
\bibfield{author}{\bibinfo{person}{Shanshan Feng}, \bibinfo{person}{Xutao Li},
  \bibinfo{person}{Yifeng Zeng}, \bibinfo{person}{Gao Cong},
  \bibinfo{person}{Yeow~Meng Chee}, {and} \bibinfo{person}{Quan Yuan}.}
  \bibinfo{year}{2015}\natexlab{}.
\newblock \showarticletitle{Personalized ranking metric embedding for next new
  POI recommendation}. In \bibinfo{booktitle}{\emph{IJCAI'15}}.
\newblock


\bibitem[\protect\citeauthoryear{Garcia-Duran, Gonzalez, Onoro-Rubio, Niepert,
  and Li}{Garcia-Duran et~al\mbox{.}}{2018}]%
        {garcia2018transrev}
\bibfield{author}{\bibinfo{person}{Alberto Garcia-Duran},
  \bibinfo{person}{Roberto Gonzalez}, \bibinfo{person}{Daniel Onoro-Rubio},
  \bibinfo{person}{Mathias Niepert}, {and} \bibinfo{person}{Hui Li}.}
  \bibinfo{year}{2018}\natexlab{}.
\newblock \showarticletitle{TransRev: Modeling Reviews as Translations from
  Users to Items}.
\newblock \bibinfo{journal}{\emph{arXiv preprint arXiv:1801.10095}}
  (\bibinfo{year}{2018}).
\newblock


\bibitem[\protect\citeauthoryear{He, Kang, and McAuley}{He
  et~al\mbox{.}}{2017a}]%
        {TransRec}
\bibfield{author}{\bibinfo{person}{Ruining He}, \bibinfo{person}{Wang{-}Cheng
  Kang}, {and} \bibinfo{person}{Julian McAuley}.}
  \bibinfo{year}{2017}\natexlab{a}.
\newblock \showarticletitle{Translation-based Recommendation}. In
  \bibinfo{booktitle}{\emph{RecSys'17}}.
\newblock


\bibitem[\protect\citeauthoryear{He and McAuley}{He and McAuley}{2016}]%
        {he2016fusing}
\bibfield{author}{\bibinfo{person}{Ruining He} {and} \bibinfo{person}{Julian
  McAuley}.} \bibinfo{year}{2016}\natexlab{}.
\newblock \showarticletitle{Fusing similarity models with markov chains for
  sparse sequential recommendation}. In \bibinfo{booktitle}{\emph{ICDM'16}}.
\newblock


\bibitem[\protect\citeauthoryear{He, Liao, Zhang, Nie, Hu, and Chua}{He
  et~al\mbox{.}}{2017b}]%
        {NeuMF}
\bibfield{author}{\bibinfo{person}{Xiangnan He}, \bibinfo{person}{Lizi Liao},
  \bibinfo{person}{Hanwang Zhang}, \bibinfo{person}{Liqiang Nie},
  \bibinfo{person}{Xia Hu}, {and} \bibinfo{person}{Tat-Seng Chua}.}
  \bibinfo{year}{2017}\natexlab{b}.
\newblock \showarticletitle{Neural collaborative filtering}. In
  \bibinfo{booktitle}{\emph{WWW'17}}.
\newblock


\bibitem[\protect\citeauthoryear{Hidasi and Karatzoglou}{Hidasi and
  Karatzoglou}{2017}]%
        {DBLP:journals/corr/HidasiK17}
\bibfield{author}{\bibinfo{person}{Bal{\'{a}}zs Hidasi} {and}
  \bibinfo{person}{Alexandros Karatzoglou}.} \bibinfo{year}{2017}\natexlab{}.
\newblock \showarticletitle{Recurrent Neural Networks with Top-k Gains for
  Session-based Recommendations}.
\newblock \bibinfo{journal}{\emph{arXiv}}  \bibinfo{volume}{abs/1706.03847}
  (\bibinfo{year}{2017}).
\newblock


\bibitem[\protect\citeauthoryear{Hidasi, Karatzoglou, Baltrunas, and
  Tikk}{Hidasi et~al\mbox{.}}{2016}]%
        {DBLP:journals/corr/HidasiKBT15}
\bibfield{author}{\bibinfo{person}{Bal{\'{a}}zs Hidasi},
  \bibinfo{person}{Alexandros Karatzoglou}, \bibinfo{person}{Linas Baltrunas},
  {and} \bibinfo{person}{Domonkos Tikk}.} \bibinfo{year}{2016}\natexlab{}.
\newblock \showarticletitle{Session-based Recommendations with Recurrent Neural
  Networks}. In \bibinfo{booktitle}{\emph{ICLR'16}}.
\newblock


\bibitem[\protect\citeauthoryear{Hsieh, Yang, Cui, Lin, Belongie, and
  Estrin}{Hsieh et~al\mbox{.}}{2017}]%
        {CML}
\bibfield{author}{\bibinfo{person}{Cheng{-}Kang Hsieh}, \bibinfo{person}{Longqi
  Yang}, \bibinfo{person}{Yin Cui}, \bibinfo{person}{Tsung{-}Yi Lin},
  \bibinfo{person}{Serge~J. Belongie}, {and} \bibinfo{person}{Deborah Estrin}.}
  \bibinfo{year}{2017}\natexlab{}.
\newblock \showarticletitle{Collaborative Metric Learning}. In
  \bibinfo{booktitle}{\emph{WWW'17}}.
\newblock


\bibitem[\protect\citeauthoryear{Hu, Koren, and Volinsky}{Hu
  et~al\mbox{.}}{2008}]%
        {WRMF}
\bibfield{author}{\bibinfo{person}{Yifan Hu}, \bibinfo{person}{Yehuda Koren},
  {and} \bibinfo{person}{Chris Volinsky}.} \bibinfo{year}{2008}\natexlab{}.
\newblock \showarticletitle{Collaborative filtering for implicit feedback
  datasets}. In \bibinfo{booktitle}{\emph{ICDM'08}}.
\newblock


\bibitem[\protect\citeauthoryear{Jacobs, Jordan, Nowlan, and Hinton}{Jacobs
  et~al\mbox{.}}{1991}]%
        {moe}
\bibfield{author}{\bibinfo{person}{Robert Jacobs}, \bibinfo{person}{Michael
  Jordan}, \bibinfo{person}{Steven Nowlan}, {and} \bibinfo{person}{Geoffrey
  Hinton}.} \bibinfo{year}{1991}\natexlab{}.
\newblock \showarticletitle{Adaptive Mixtures of Local Experts}.
\newblock \bibinfo{journal}{\emph{Neural Computation}} (\bibinfo{year}{1991}).
\newblock


\bibitem[\protect\citeauthoryear{Jing and Smola}{Jing and Smola}{2017}]%
        {DBLP:conf/wsdm/JingS17}
\bibfield{author}{\bibinfo{person}{How Jing} {and}
  \bibinfo{person}{Alexander~J. Smola}.} \bibinfo{year}{2017}\natexlab{}.
\newblock \showarticletitle{Neural Survival Recommender}. In
  \bibinfo{booktitle}{\emph{WSDM'17}}.
\newblock


\bibitem[\protect\citeauthoryear{Kabbur, Ning, and Karypis}{Kabbur
  et~al\mbox{.}}{2013}]%
        {kabbur2013fism}
\bibfield{author}{\bibinfo{person}{Santosh Kabbur}, \bibinfo{person}{Xia Ning},
  {and} \bibinfo{person}{George Karypis}.} \bibinfo{year}{2013}\natexlab{}.
\newblock \showarticletitle{{FISM:} factored item similarity models for top-n
  recommender systems}. In \bibinfo{booktitle}{\emph{KDD'13}}.
\newblock


\bibitem[\protect\citeauthoryear{Kang, Fang, Wang, and McAuley}{Kang
  et~al\mbox{.}}{2017}]%
        {DBLP:conf/icdm/KangFWM17}
\bibfield{author}{\bibinfo{person}{Wang{-}Cheng Kang}, \bibinfo{person}{Chen
  Fang}, \bibinfo{person}{Zhaowen Wang}, {and} \bibinfo{person}{Julian
  McAuley}.} \bibinfo{year}{2017}\natexlab{}.
\newblock \showarticletitle{Visually-Aware Fashion Recommendation and Design
  with Generative Image Models}. In \bibinfo{booktitle}{\emph{ICDM'17}}.
\newblock


\bibitem[\protect\citeauthoryear{Kim, Park, Oh, Lee, and Yu}{Kim
  et~al\mbox{.}}{2016}]%
        {DBLP:conf/recsys/KimPOLY16}
\bibfield{author}{\bibinfo{person}{Dong~Hyun Kim}, \bibinfo{person}{Chanyoung
  Park}, \bibinfo{person}{Jinoh Oh}, \bibinfo{person}{Sungyoung Lee}, {and}
  \bibinfo{person}{Hwanjo Yu}.} \bibinfo{year}{2016}\natexlab{}.
\newblock \showarticletitle{Convolutional Matrix Factorization for Document
  Context-Aware Recommendation}. In \bibinfo{booktitle}{\emph{RecSys'16}}.
\newblock


\bibitem[\protect\citeauthoryear{Kingma and Ba}{Kingma and Ba}{2014}]%
        {adam}
\bibfield{author}{\bibinfo{person}{Diederik~P Kingma} {and}
  \bibinfo{person}{Jimmy Ba}.} \bibinfo{year}{2014}\natexlab{}.
\newblock \showarticletitle{Adam: A method for stochastic optimization}.
\newblock \bibinfo{journal}{\emph{arXiv preprint arXiv:1412.6980}}
  (\bibinfo{year}{2014}).
\newblock


\bibitem[\protect\citeauthoryear{Koren}{Koren}{2008}]%
        {koren2008factorization}
\bibfield{author}{\bibinfo{person}{Yehuda Koren}.}
  \bibinfo{year}{2008}\natexlab{}.
\newblock \showarticletitle{Factorization meets the neighborhood: a
  multifaceted collaborative filtering model}. In
  \bibinfo{booktitle}{\emph{KDD'08}}.
\newblock


\bibitem[\protect\citeauthoryear{Koren}{Koren}{2010}]%
        {koren2010temporal}
\bibfield{author}{\bibinfo{person}{Yehuda Koren}.}
  \bibinfo{year}{2010}\natexlab{}.
\newblock \showarticletitle{Collaborative filtering with temporal dynamics}.
\newblock \bibinfo{journal}{\emph{Commun. ACM}} (\bibinfo{year}{2010}).
\newblock


\bibitem[\protect\citeauthoryear{Koren, Bell, and Volinsky}{Koren
  et~al\mbox{.}}{2009}]%
        {koren2009matrix}
\bibfield{author}{\bibinfo{person}{Yehuda Koren}, \bibinfo{person}{Robert
  Bell}, {and} \bibinfo{person}{Chris Volinsky}.}
  \bibinfo{year}{2009}\natexlab{}.
\newblock \showarticletitle{Matrix Factorization techniques for recommender
  systems}.
\newblock \bibinfo{journal}{\emph{Computer}} (\bibinfo{year}{2009}).
\newblock


\bibitem[\protect\citeauthoryear{Li, Ren, Chen, Ren, Lian, and Ma}{Li
  et~al\mbox{.}}{2017}]%
        {DBLP:conf/cikm/LiRCRLM17}
\bibfield{author}{\bibinfo{person}{Jing Li}, \bibinfo{person}{Pengjie Ren},
  \bibinfo{person}{Zhumin Chen}, \bibinfo{person}{Zhaochun Ren},
  \bibinfo{person}{Tao Lian}, {and} \bibinfo{person}{Jun Ma}.}
  \bibinfo{year}{2017}\natexlab{}.
\newblock \showarticletitle{Neural Attentive Session-based Recommendation}. In
  \bibinfo{booktitle}{\emph{CIKM'17}}.
\newblock


\bibitem[\protect\citeauthoryear{Lin, Liu, Sun, Liu, and Zhu}{Lin
  et~al\mbox{.}}{2015}]%
        {TransR}
\bibfield{author}{\bibinfo{person}{Yankai Lin}, \bibinfo{person}{Zhiyuan Liu},
  \bibinfo{person}{Maosong Sun}, \bibinfo{person}{Yang Liu}, {and}
  \bibinfo{person}{Xuan Zhu}.} \bibinfo{year}{2015}\natexlab{}.
\newblock \showarticletitle{Learning Entity and Relation Embeddings for
  Knowledge Graph Completion}. In \bibinfo{booktitle}{\emph{AAAI'15}}.
\newblock


\bibitem[\protect\citeauthoryear{Liu, Fu, Yao, and Xiong}{Liu
  et~al\mbox{.}}{2013}]%
        {liu2013learning}
\bibfield{author}{\bibinfo{person}{Bin Liu}, \bibinfo{person}{Yanjie Fu},
  \bibinfo{person}{Zijun Yao}, {and} \bibinfo{person}{Hui Xiong}.}
  \bibinfo{year}{2013}\natexlab{}.
\newblock \showarticletitle{Learning geographical preferences for
  point-of-interest recommendation}. In \bibinfo{booktitle}{\emph{KDD'13}}.
\newblock


\bibitem[\protect\citeauthoryear{McAuley, Pandey, and Leskovec}{McAuley
  et~al\mbox{.}}{2015a}]%
        {mcauley2015inferring}
\bibfield{author}{\bibinfo{person}{Julian McAuley}, \bibinfo{person}{Rahul
  Pandey}, {and} \bibinfo{person}{Jure Leskovec}.}
  \bibinfo{year}{2015}\natexlab{a}.
\newblock \showarticletitle{Inferring networks of substitutable and
  complementary products}. In \bibinfo{booktitle}{\emph{KDD'15}}.
\newblock


\bibitem[\protect\citeauthoryear{McAuley, Targett, Shi, and van~den
  Hengel}{McAuley et~al\mbox{.}}{2015b}]%
        {mcauley2015image}
\bibfield{author}{\bibinfo{person}{Julian McAuley},
  \bibinfo{person}{Christopher Targett}, \bibinfo{person}{Qinfeng Shi}, {and}
  \bibinfo{person}{Anton van~den Hengel}.} \bibinfo{year}{2015}\natexlab{b}.
\newblock \showarticletitle{Image-based recommendations on styles and
  substitutes}. In \bibinfo{booktitle}{\emph{SIGIR'15}}.
\newblock


\bibitem[\protect\citeauthoryear{Park, Kim, Oh, and Yu}{Park
  et~al\mbox{.}}{2017}]%
        {MCF}
\bibfield{author}{\bibinfo{person}{Chanyoung Park}, \bibinfo{person}{Dong~Hyun
  Kim}, \bibinfo{person}{Jinoh Oh}, {and} \bibinfo{person}{Hwanjo Yu}.}
  \bibinfo{year}{2017}\natexlab{}.
\newblock \showarticletitle{Do "Also-Viewed" Products Help User Rating
  Prediction?}. In \bibinfo{booktitle}{\emph{WWW'17}}.
\newblock


\bibitem[\protect\citeauthoryear{Pathak, Gupta, and McAuley}{Pathak
  et~al\mbox{.}}{2017}]%
        {bundle}
\bibfield{author}{\bibinfo{person}{Apurva Pathak}, \bibinfo{person}{Kshitiz
  Gupta}, {and} \bibinfo{person}{Julian McAuley}.}
  \bibinfo{year}{2017}\natexlab{}.
\newblock \showarticletitle{Generating and Personalizing Bundle Recommendations
  on \emph{Steam}}. In \bibinfo{booktitle}{\emph{SIGIR'17}}.
\newblock


\bibitem[\protect\citeauthoryear{Pei, Yang, Sun, Zhang, Bozzon, and Tax}{Pei
  et~al\mbox{.}}{2017}]%
        {pei2017interacting}
\bibfield{author}{\bibinfo{person}{Wenjie Pei}, \bibinfo{person}{Jie Yang},
  \bibinfo{person}{Zhu Sun}, \bibinfo{person}{Jie Zhang},
  \bibinfo{person}{Alessandro Bozzon}, {and} \bibinfo{person}{David~MJ Tax}.}
  \bibinfo{year}{2017}\natexlab{}.
\newblock \showarticletitle{Interacting Attention-gated Recurrent Networks for
  Recommendation}. In \bibinfo{booktitle}{\emph{CIKM'17}}.
\newblock


\bibitem[\protect\citeauthoryear{Rendle, Freudenthaler, Gantner, and
  Schmidt-Thieme}{Rendle et~al\mbox{.}}{2009}]%
        {rendle2009bpr}
\bibfield{author}{\bibinfo{person}{Steffen Rendle}, \bibinfo{person}{Christoph
  Freudenthaler}, \bibinfo{person}{Zeno Gantner}, {and} \bibinfo{person}{Lars
  Schmidt-Thieme}.} \bibinfo{year}{2009}\natexlab{}.
\newblock \showarticletitle{{BPR:} Bayesian personalized ranking from implicit
  feedback}. In \bibinfo{booktitle}{\emph{UAI'09}}.
\newblock


\bibitem[\protect\citeauthoryear{Rendle, Freudenthaler, and
  Schmidt-Thieme}{Rendle et~al\mbox{.}}{2010}]%
        {rendle2010fpmc}
\bibfield{author}{\bibinfo{person}{Steffen Rendle}, \bibinfo{person}{Christoph
  Freudenthaler}, {and} \bibinfo{person}{Lars Schmidt-Thieme}.}
  \bibinfo{year}{2010}\natexlab{}.
\newblock \showarticletitle{Factorizing personalized markov chains for
  next-basket recommendation}. In \bibinfo{booktitle}{\emph{WWW'10}}.
\newblock


\bibitem[\protect\citeauthoryear{Sedhain, Menon, Sanner, and Xie}{Sedhain
  et~al\mbox{.}}{2015}]%
        {sedhain2015autorec}
\bibfield{author}{\bibinfo{person}{Suvash Sedhain},
  \bibinfo{person}{Aditya~Krishna Menon}, \bibinfo{person}{Scott Sanner}, {and}
  \bibinfo{person}{Lexing Xie}.} \bibinfo{year}{2015}\natexlab{}.
\newblock \showarticletitle{Autorec: Autoencoders meet collaborative
  filtering}. In \bibinfo{booktitle}{\emph{WWW'15}}.
\newblock


\bibitem[\protect\citeauthoryear{Shu, Wang, Tang, Wang, and Liu}{Shu
  et~al\mbox{.}}{2018}]%
        {shu2018crossfire}
\bibfield{author}{\bibinfo{person}{Kai Shu}, \bibinfo{person}{Suhang Wang},
  \bibinfo{person}{Jiliang Tang}, \bibinfo{person}{Yilin Wang}, {and}
  \bibinfo{person}{Huan Liu}.} \bibinfo{year}{2018}\natexlab{}.
\newblock \showarticletitle{Crossfire: Cross media joint friend and item
  recommendations}. In \bibinfo{booktitle}{\emph{WSDM'18}}.
\newblock


\bibitem[\protect\citeauthoryear{Smirnova and Vasile}{Smirnova and
  Vasile}{2017}]%
        {DBLP:conf/recsys/SmirnovaV17}
\bibfield{author}{\bibinfo{person}{Elena Smirnova} {and}
  \bibinfo{person}{Flavian Vasile}.} \bibinfo{year}{2017}\natexlab{}.
\newblock \showarticletitle{Contextual Sequence Modeling for Recommendation
  with Recurrent Neural Networks}. In
  \bibinfo{booktitle}{\emph{DLRS@RecSys'17}}.
\newblock


\bibitem[\protect\citeauthoryear{Tang and Wang}{Tang and Wang}{2018}]%
        {DBLP:conf/wsdm/TangW18}
\bibfield{author}{\bibinfo{person}{Jiaxi Tang} {and} \bibinfo{person}{Ke
  Wang}.} \bibinfo{year}{2018}\natexlab{}.
\newblock \showarticletitle{Personalized Top-N Sequential Recommendation via
  Convolutional Sequence Embedding}. In \bibinfo{booktitle}{\emph{WSDM'18}}.
\newblock


\bibitem[\protect\citeauthoryear{Tay, Tuan, and Hui}{Tay et~al\mbox{.}}{2018}]%
        {tay2018latent}
\bibfield{author}{\bibinfo{person}{Yi Tay}, \bibinfo{person}{Luu~Anh Tuan},
  {and} \bibinfo{person}{Siu~Cheung Hui}.} \bibinfo{year}{2018}\natexlab{}.
\newblock \showarticletitle{Latent Relational Metric Learning via Memory-based
  A ention for Collaborative Ranking}. In \bibinfo{booktitle}{\emph{WWW'18}}.
\newblock


\bibitem[\protect\citeauthoryear{Tuan and Phuong}{Tuan and Phuong}{2017}]%
        {DBLP:conf/recsys/TuanP17}
\bibfield{author}{\bibinfo{person}{Trinh~Xuan Tuan} {and}
  \bibinfo{person}{Tu~Minh Phuong}.} \bibinfo{year}{2017}\natexlab{}.
\newblock \showarticletitle{3D Convolutional Networks for Session-based
  Recommendation with Content Features}. In
  \bibinfo{booktitle}{\emph{RecSys'17}}.
\newblock


\bibitem[\protect\citeauthoryear{Wang, Wang, and Yeung}{Wang
  et~al\mbox{.}}{2015}]%
        {DBLP:conf/kdd/WangWY15}
\bibfield{author}{\bibinfo{person}{Hao Wang}, \bibinfo{person}{Naiyan Wang},
  {and} \bibinfo{person}{Dit{-}Yan Yeung}.} \bibinfo{year}{2015}\natexlab{}.
\newblock \showarticletitle{Collaborative Deep Learning for Recommender
  Systems}. In \bibinfo{booktitle}{\emph{KDD'15}}.
\newblock


\bibitem[\protect\citeauthoryear{Wang, Wang, Tang, Shu, Ranganath, and
  Liu}{Wang et~al\mbox{.}}{2017}]%
        {wang2017your}
\bibfield{author}{\bibinfo{person}{Suhang Wang}, \bibinfo{person}{Yilin Wang},
  \bibinfo{person}{Jiliang Tang}, \bibinfo{person}{Kai Shu},
  \bibinfo{person}{Suhas Ranganath}, {and} \bibinfo{person}{Huan Liu}.}
  \bibinfo{year}{2017}\natexlab{}.
\newblock \showarticletitle{What your images reveal: Exploiting visual contents
  for point-of-interest recommendation}. In \bibinfo{booktitle}{\emph{WWW'17}}.
\newblock


\bibitem[\protect\citeauthoryear{Wu, Ge, Liu, Chen, Hong, Du, and Wang}{Wu
  et~al\mbox{.}}{2017}]%
        {wu2017modeling}
\bibfield{author}{\bibinfo{person}{Le Wu}, \bibinfo{person}{Yong Ge},
  \bibinfo{person}{Qi Liu}, \bibinfo{person}{Enhong Chen},
  \bibinfo{person}{Richang Hong}, \bibinfo{person}{Junping Du}, {and}
  \bibinfo{person}{Meng Wang}.} \bibinfo{year}{2017}\natexlab{}.
\newblock \showarticletitle{Modeling the evolution of users' preferences and
  social links in social networking services}.
\newblock \bibinfo{journal}{\emph{IEEE TKDE}} (\bibinfo{year}{2017}).
\newblock


\bibitem[\protect\citeauthoryear{Yang, Bai, Zhang, Yuan, and Han}{Yang
  et~al\mbox{.}}{2017}]%
        {PACE}
\bibfield{author}{\bibinfo{person}{Carl Yang}, \bibinfo{person}{Lanxiao Bai},
  \bibinfo{person}{Chao Zhang}, \bibinfo{person}{Quan Yuan}, {and}
  \bibinfo{person}{Jiawei Han}.} \bibinfo{year}{2017}\natexlab{}.
\newblock \showarticletitle{Bridging Collaborative Filtering and
  Semi-Supervised Learning: {A} Neural Approach for {POI} Recommendation}. In
  \bibinfo{booktitle}{\emph{KDD'17}}.
\newblock


\bibitem[\protect\citeauthoryear{Yi, Hong, Zhong, Liu, and Rajan}{Yi
  et~al\mbox{.}}{2014}]%
        {yi2014beyond}
\bibfield{author}{\bibinfo{person}{Xing Yi}, \bibinfo{person}{Liangjie Hong},
  \bibinfo{person}{Erheng Zhong}, \bibinfo{person}{Nanthan~Nan Liu}, {and}
  \bibinfo{person}{Suju Rajan}.} \bibinfo{year}{2014}\natexlab{}.
\newblock \showarticletitle{Beyond clicks: dwell time for personalization}. In
  \bibinfo{booktitle}{\emph{RecSys'14}}.
\newblock


\bibitem[\protect\citeauthoryear{Yu, Ren, Sun, Gu, Sturt, Khandelwal, Norick,
  and Han}{Yu et~al\mbox{.}}{2014}]%
        {yu2014personalized}
\bibfield{author}{\bibinfo{person}{Xiao Yu}, \bibinfo{person}{Xiang Ren},
  \bibinfo{person}{Yizhou Sun}, \bibinfo{person}{Quanquan Gu},
  \bibinfo{person}{Bradley Sturt}, \bibinfo{person}{Urvashi Khandelwal},
  \bibinfo{person}{Brandon Norick}, {and} \bibinfo{person}{Jiawei Han}.}
  \bibinfo{year}{2014}\natexlab{}.
\newblock \showarticletitle{Personalized entity recommendation: A heterogeneous
  information network approach}. In \bibinfo{booktitle}{\emph{WSDM'14}}.
\newblock


\bibitem[\protect\citeauthoryear{Zhang, Yuan, Lian, Xie, and Ma}{Zhang
  et~al\mbox{.}}{2016}]%
        {CKE}
\bibfield{author}{\bibinfo{person}{Fuzheng Zhang},
  \bibinfo{person}{Nicholas~Jing Yuan}, \bibinfo{person}{Defu Lian},
  \bibinfo{person}{Xing Xie}, {and} \bibinfo{person}{Wei{-}Ying Ma}.}
  \bibinfo{year}{2016}\natexlab{}.
\newblock \showarticletitle{Collaborative Knowledge Base Embedding for
  Recommender Systems}. In \bibinfo{booktitle}{\emph{KDD'16}}.
\newblock


\bibitem[\protect\citeauthoryear{Zhang, Yao, and Sun}{Zhang
  et~al\mbox{.}}{2017}]%
        {DBLP:journals/corr/ZhangYS17aa}
\bibfield{author}{\bibinfo{person}{Shuai Zhang}, \bibinfo{person}{Lina Yao},
  {and} \bibinfo{person}{Aixin Sun}.} \bibinfo{year}{2017}\natexlab{}.
\newblock \showarticletitle{Deep Learning based Recommender System: {A} Survey
  and New Perspectives}.
\newblock \bibinfo{journal}{\emph{arXiv}}  \bibinfo{volume}{abs/1707.07435}
  (\bibinfo{year}{2017}).
\newblock


\bibitem[\protect\citeauthoryear{Zhao, Yao, Li, Song, and Lee}{Zhao
  et~al\mbox{.}}{2017}]%
        {zhao2017meta}
\bibfield{author}{\bibinfo{person}{Huan Zhao}, \bibinfo{person}{Quanming Yao},
  \bibinfo{person}{Jianda Li}, \bibinfo{person}{Yangqiu Song}, {and}
  \bibinfo{person}{Dik~Lun Lee}.} \bibinfo{year}{2017}\natexlab{}.
\newblock \showarticletitle{Meta-graph based recommendation fusion over
  heterogeneous information networks}. In \bibinfo{booktitle}{\emph{KDD'17}}.
\newblock


\bibitem[\protect\citeauthoryear{Zhao, McAuley, and King}{Zhao
  et~al\mbox{.}}{2014}]%
        {zhao2014leveraging}
\bibfield{author}{\bibinfo{person}{Tong Zhao}, \bibinfo{person}{Julian
  McAuley}, {and} \bibinfo{person}{Irwin King}.}
  \bibinfo{year}{2014}\natexlab{}.
\newblock \showarticletitle{Leveraging social connections to improve
  personalized ranking for collaborative filtering}. In
  \bibinfo{booktitle}{\emph{CIKM'14}}.
\newblock


\end{thebibliography}

\end{document}